\documentclass[aip,amsmath,amssymb,reprint]{revtex4-1}

\usepackage[T1]{fontenc}
\usepackage[utf8]{inputenc}
\usepackage[english]{babel}
\usepackage{graphicx}
\usepackage{dcolumn}
\usepackage{bm}

\usepackage[utf8]{inputenc}
\usepackage[T1]{fontenc}
\usepackage{mathptmx}
\usepackage{etoolbox}

\makeatletter
\def\@email#1#2{%
 \endgroup
 \patchcmd{\titleblock@produce}
  {\frontmatter@RRAPformat}
  {\frontmatter@RRAPformat{\produce@RRAP{*#1\href{mailto:#2}{#2}}}\frontmatter@RRAPformat}
  {}{}
}%
\makeatother
\usepackage{amsfonts,amsbsy,amssymb,amsmath,graphicx}

\graphicspath{{figures/}}
\usepackage[linktoc=all,colorlinks=true,citecolor=blue,linkcolor=blue]{hyperref}
\newcommand{\nico}[1]{\textcolor{black}{#1}}
\newcommand{\niconew}[1]{\textcolor{black}{#1}}
\begin{document}	
\preprint{AIP/123-QED}

\title[]{Building transport models from baroclinic wave experimental data}
\author{M. Agaoglou}
\affiliation{Departamento de Matem{\'a}tica Aplicada a la Ingenier{\'i}a Industrial, Escuela T{\'e}cnica Superior de Ingenieros Industriales, Universidad Polit{\'e}cnica de Madrid, 28006 Madrid, Spain.
}%
\author{V. J. Garc\'ia-Garrido}%
\affiliation{Departamento de F\'isica y Matem\'aticas, Facultad de Ciencias, Universidad de Alcal\'a, 28805 Alcal\'a de Henares, Madrid, Spain. 
}%
\author{U. Harlander}%
\affiliation{Department of Aerodynamics and Fluid Mechanics,
Brandenburg University of Technology Cottbus, D-03046
Cottbus, Germany.
}%

\author{A. M. Mancho}
\altaffiliation[Corresponding author:]{a.m.mancho@icmat.es}
\affiliation{Instituto de Ciencias Matem\'aticas. Consejo Superior de Investigaciones Cient\'ificas, 28049 Madrid, Spain. 
}

\email{a.m.mancho@icmat.es}
 
\date{\today}

\title{Building transport models from baroclinic wave experimental data} 
	
 	
\begin{abstract}
\nico{In this paper we study baroclinic waves both from the experimental and the theoretical perspective. We obtain data from a rotating annulus experiment capable of producing a series of baroclinic eddies similar to those found in the mid-latitude atmosphere. We analyze the experimental outputs using two methods. First, we apply a technique that involves filtering data using Empirical Orthogonal Function (EOF) analysis, which is applied to both velocity and surface temperature fields. The second method relies on the construction of a simple kinematic model based on key parameters derived from the experimental data. To analyze eddy-driven fluid transport, we apply the method of Lagrangian descriptors to the underlying velocity field, revealing the attracting material curves that act as transport barriers in the system. These structures effectively capture the essential characteristics of the baroclinic flow and the associated transport phenomena. Our results show that these barriers are in good agreement with the transport patterns observed in the rotating annulus experiment. In particular, we observe that the structures obtained from the kinematic model, or the one derived in terms of filtered velocities, perform well in this regard. }

\end{abstract}

\maketitle	



\section{Introduction}
\label{sec:intro}
Baroclinic instability is a significant contributor to the variability of mid-latitude weather. Baroclinic waves develop in a nearly geostrophic flow where the wind above the Ekman layer is blowing parallel to the temperature isotherms with a nonzero temperature gradient across the flow. This gradient produces a vertical shear in the geostrophic flow that becomes unstable when a critical threshold of the temperature gradient is exceeded. Then a small initial perturbation superimposed on the flow will grow, and this situation is known as a state of \emph{baroclinic instability} \cite{Pedlosky:89}. In the atmosphere, this instability often occurs in the mid-latitudes, between 30 and 60 degrees latitude, where the polar air mass meets the warmer subtropical air mass. This instability is responsible for most of the day-to-day variability in weather in mid-latitude regions and plays a fundamental role also for non-terrestrial planetary atmospheres \cite{Read_etal:2020}. The wave patterns of the mid-latitude weather typically have a frontal system with warm and cold fronts and are associated with rain, snow, wind, and changes in temperature.
  
To understand this instability and how the atmospheric circulation transports heat from equatorial to polar latitudes in the mid-latitude dynamics of the Earth's atmosphere, an elegant laboratory experiment was designed in the 1950s \cite{Fowlis_and_Hide:65}. This experiment consists of a differentially heated rotating annulus where a jet is indirectly forced by a radial temperature gradient. The jet is baroclinically unstable and has many features in common with the jets in planetary atmospheres \cite{read_sommeria_young_2019} since \emph{baroclinic waves} dominate the experimental flow. The experiment contains a cylinder with a cooled inner and heated outer wall placed on a rotating platform. Although the cylindrical geometry differs from the sphere, the differential heating mimics the Earth's temperature difference between the tropics and the poles. Different wave regimes have been identified in this setting depending on the strength of the heating and the rate of rotation. These  can be classified by pro-grade propagating waves of different wavenumbers and quasi-chaotic regimes where waves and small-scale vortices coexist. Over the years many aspects of atmospheric flows have been covered using the rotating annulus experiment. During the first years linear baroclinic instability was identified as a key process of large-scale dynamics \cite{Fowlis_and_Hide:65}. Moreover, nonlinear regime transitions and bifurcations to chaotic flows have been analyzed \cite{Morita:90}. Of interest was also the nonlinear saturation of baroclinic waves \cite{Hart:81}, wave-mean flow, and wave-wave interactions \cite{Frueh_and_Read:97}. Also vascillations and low-frequency variability \cite{Lindzen_etal:81} have been investigated intensively. Recently, secondary instabilities of the baroclinic jet and the generation of internal gravity waves have been studied \cite{Rodda_and_Harlander:20}. The studies have been done with a flat bottom (f-plane) or with a bottom cone leading to a linear increase of fluid depth towards the outer cylinder \cite{Vincze_etal:2014}. The latter is called \emph{topographic $\beta$-plane} and it simulates the Earth's changing Coriolis parameter with latitude ($\beta$-plane). A review of findings from the differentially heated rotating annulus experiment summarizing many aspects mentioned above can be found in \cite{Read_etal:2014}. 

When baroclinic waves develop, either  in the atmosphere, the  ocean, or  lab experiments, they give rise to  mixing and heat transport phenomena across the jet stream. Experimental studies have been conducted to investigate the intricate interactions between the baroclinic waves, mixing, and the transport of heat or  other tracers across these jets. 
For instance, \cite{Sommeria1} injected dye in the fluid of a rotating annulus to show that the jet acts as a barrier to tracer transport, similar to the southern stratospheric polar night jet may act as a barrier to the transport of trace gases from lower to polar latitudes.  They reveal a striking barrier to mixing across the jet and such barriers prohibit trajectories from crossing the meandering jet flow. In \cite{Keane_etal14}, stirring properties of the rotating annulus flow were investigated showing that finite scale Lyapunov exponents correctly identify barriers between different flow regimes. These studies connect well to others like \cite{bernard,alvaro10,alvaro1,alvaro2,Garcia2017,Curbelo,Curbelo_partI,Curbelo_partII} exploring   
transport and mixing phenomena occurring in the antarctic polar vortex in the southern stratosphere using reanalysis data. 
Additional experiments on mixing on azimuthal jets in a rapidly rotating annular tank have also been performed in \cite{behringer1991}.

This paper aims to study this phenomenon further, both from the experimental and theoretical point of view. The experimental setup, as is usually done in this field,  \nico{consists of a rotating annulus experiment with differential heating in the radial direction achieved by cooling the vertical interior and heating the vertical exterior wall. Recently, some variations with respect to the thermal forcing has been discussed \cite{Scolan_etal:2017, Svarnakar:2023} that show some interesting new phenomena.} However, as a distinctive feature in our case, the measuring instruments will simultaneously collect the temperature of the surface and the horizontal velocity in a horizontal plane below it.  
In order to examine the experimental results from a theoretical perspective, we will adopt simplified models that effectively capture the essential characteristics of baroclinic instability and the associated transport phenomena. Rather than solving the complete set of governing equations, these simplified models provide a framework to analyze and understand the key features of the system. By employing such models, we can focus on the fundamental dynamics and gain insights into the mechanisms underlying baroclinic instability and the transport processes occurring across it.
Among the simplified models employed in the geophysical fluid dynamics community, kinematic models have a rich historical background and are frequently employed to investigate Lagrangian transport and exchange across jets. These models offer a straightforward approach and have been widely utilized in numerous studies \cite{AGP,Castillo,Samelson,Wiggins,Sommeria1,Mancho2,Sommeria2}. Additionally, with the large availability of geophysical and experimental data, models based on machine learning techniques have recently become prevalent. These models use techniques such as principal component analysis (PCA), also referred to as Karhunen-Lo\`{e}ve decomposition or Empirical Orthogonal Function (EOF) analysis in the context of atmospheric sciences, dynamic mode decomposition, and the Koopman operator to reconstruct spatiotemporal signals of complex systems \cite{uhl,kut,mezic1,mezic2}. This paper will explore model building both from a kinematic perspective and an EOF-based approach. We expect that the use of these two methodologies will allow a deeper analysis of the system.

The paper is outlined as follows. In Section \ref{sec:experimemtal_setup} we describe the experimental setup and the type of data collected to perform the analysis. Section \ref{sec: models} is devoted to describing the procedures to obtain the models from the data, including a kinematic model and the EOF analysis. 
In Sec. \ref{sec: results} we present the results that compare the performance between models and experiments and the discussion. The comparison is implemented in terms of the analysis of the transport produced by the models, which is  supported by dynamical systems ideas also discussed in this section.  
We finish this work by summarizing the conclusions in Sec. \ref{conclusions}.

\begin{table}[]
    \centering
    \begin{tabular}{|c|c|c|c|}
       \hline
       \textbf{geometry}  & & &   \\
       \hline
          & inner radius   & a (mm) & 45\\
          & outer radius   & b (mm) & 120\\
          & gap width      & b-a (mm) & 75\\
          & fluid depth & d (mm) & 135\\
          
    \hline
    \textbf{exp. parameters} & & & \\
    \hline
    & temperature difference & $\Delta T$ (K) & 4.0\\
    & revolution speed & $\Omega$ (rpm) & 4.8\\
    \hline
    \textbf{fluid properties} & & & \\
    \hline
    & density & $\rho$ ($kg m^{-3}$) & 997\\
    & kin. viscosity & $\nu$ ($m^2 s^{-1}$) & 1.004 $\times 10^{-6}$ \\
    & therm. conductivity & $\kappa$ ($m^2 s^{-1}$) & 0.1434 $\times 10^{-6}$ \\
    & exp. coefficient & $\alpha$ (1/K) & 0.207 $\times 10^{-3}$ \\
    & Prandtl number & Pr & 7\\
    \hline
    \end{tabular}
    \caption{Parameters of the laboratory experiment. Surface temperature and horizontal velocity 1 cm below the surface were measured simultaneously. The sampling frequency was $0.16$ Hz and 542 pictures were taken ($\approx 56.5$ minutes of measurement). For the laboratory setup, see Fig. \ref{setup}.}
    \label{tab:my_label}
\end{table}

\section{The experimental setup and data acquisition}
\label{sec:experimemtal_setup}

\begin{figure}[htbp]
	\begin{center}
		\includegraphics[scale=0.87]{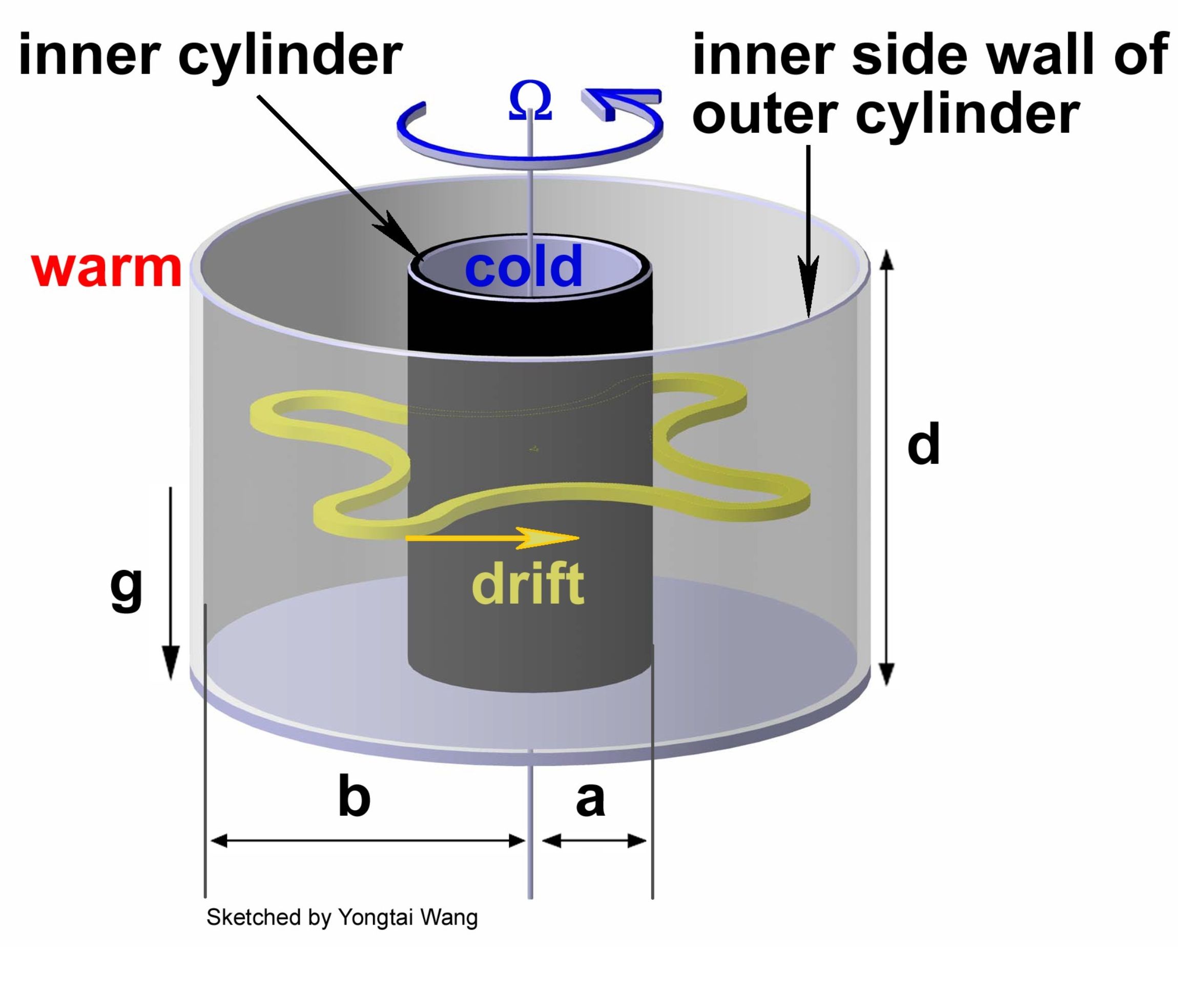}
		\includegraphics[scale=0.03]{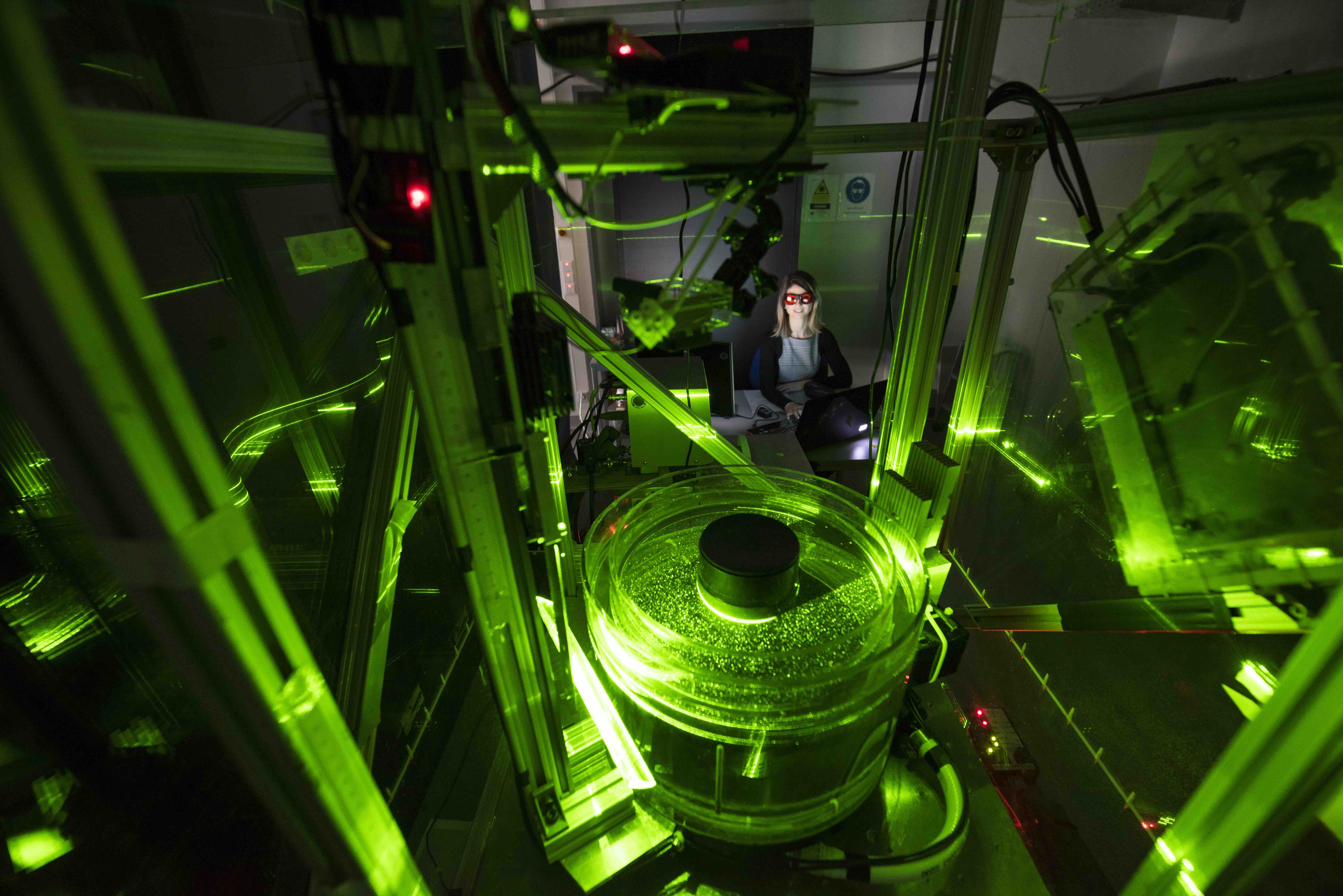}
	\end{center}
	\caption{Set-up of the experiment. (a) Sketch of the rotating baroclinic wave tank with the heated outer wall and the cooled inner cylinder. In the unstable flow regime a baroclinic wave forms that mainly drifts with the mean flow in the prograde direction. The drift is slightly faster than the angular velocity $\Omega$ of the tank. (b) The experimental unit with the tank in the foreground and the computer control unit in the background. The tank is mounted on a rotating platform and is protected in a perspex box to minimize the influence of air movements in the room. The outer part is heated by a heating coil, the inner part is cooled via a thermostat.}\label{setup}
\end{figure}

The tank consists of three concentric cylinders mounted on a turntable, see Fig. \ref{setup}. While the inner cylinder is made of anodized aluminum, the middle and outer one is made of borosilicate glass. \niconew{The inner cylinder shown in Fig. \ref{setup} (top) is cooled by an externally positioned lab thermostat of type ``Julabo F25''. Lab thermostats can not only control the temperature, they can also cool or heat the fluid. Technically, the cooling is done in a similar way than in a compression refrigerator but the technical details depend on the manufacturer. The thermostat receives feedback from a temperature sensor mounted in the inner cylinder. In a lab thermostat, a proportional–integral–derivative (PID) controller, widely used in industrial control systems requiring continuously modulated 
control takes care of keeping the temperature constant.} A pump with a maximum pump capacity of $20$ l/min is pumping the cold fluid from the thermostat into the inner cylinder and another pump with a maximum pump capacity of $14$ l/min extracts the water from the inner cylinder and transports it back to the thermostat. The tubes are made of copper and PVC. The outer annulus of the experiment visible in Fig. \ref{setup} (bottom) is heated by a heating coil that is mounted at the bottom. \niconew{The temperature of the heating coil is regulated by the voltage applied.} A pump with a maximum pump capacity of $20$ l/min pumps the water through the outer heating chamber to avoid temperature inhomogeneities. Temperature sensors in the inner and outer bath (Pt 100, accuracy $\pm 0.1$ K) measure the temperature difference $\Delta T$.
The experiment has a flat bottom and a free surface. 
\nico{To avoid heat flux at the bottom, the aluminum bottom is covered by a $5$ mm thick layer of polyoxymethylene (POM). Note that there is a nonzero heat flux and evaporation at the free surface. With respect to the baroclinic wave stability diagram by \cite{Fowlis_and_Hide:65}, this effect is rather small since the surface area of the annulus is small. However, as described by \cite{Fein:1973}, the drift speed of the waves is higher for experiments with a free surface. For bigger systems \cite{Sukha:2023, Rodda_etal:2020, Harlander_etal_atmos23} it can be expected that surface effects become stronger due to a larger surface area and might become an important issue. Note further that surface waves are not excited at the free upper boundary. The surface has a parabolic radial profile due to the tank’s rotation. However, the deflection is very small. For $\Omega=4.8$ rpm the difference in height at the inner and outer cylinder is just $0.14$ mm. That is, the upper free surface is virtually flat.}
The relevant parameters for the experiment can be found in Table \ref{tab:my_label}. As a working fluid deionized water is used. 

\begin{figure*}[htbp]
\includegraphics[scale=.18]{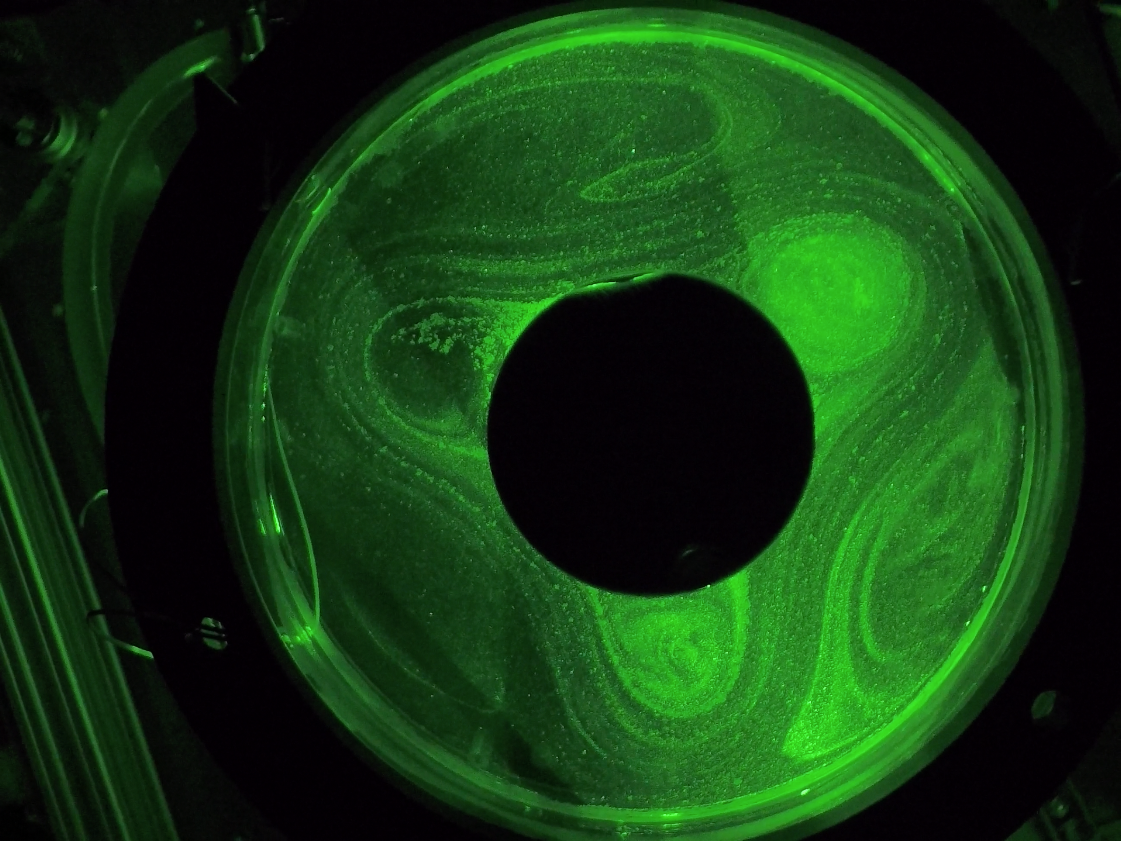} 
\includegraphics[scale=.18]{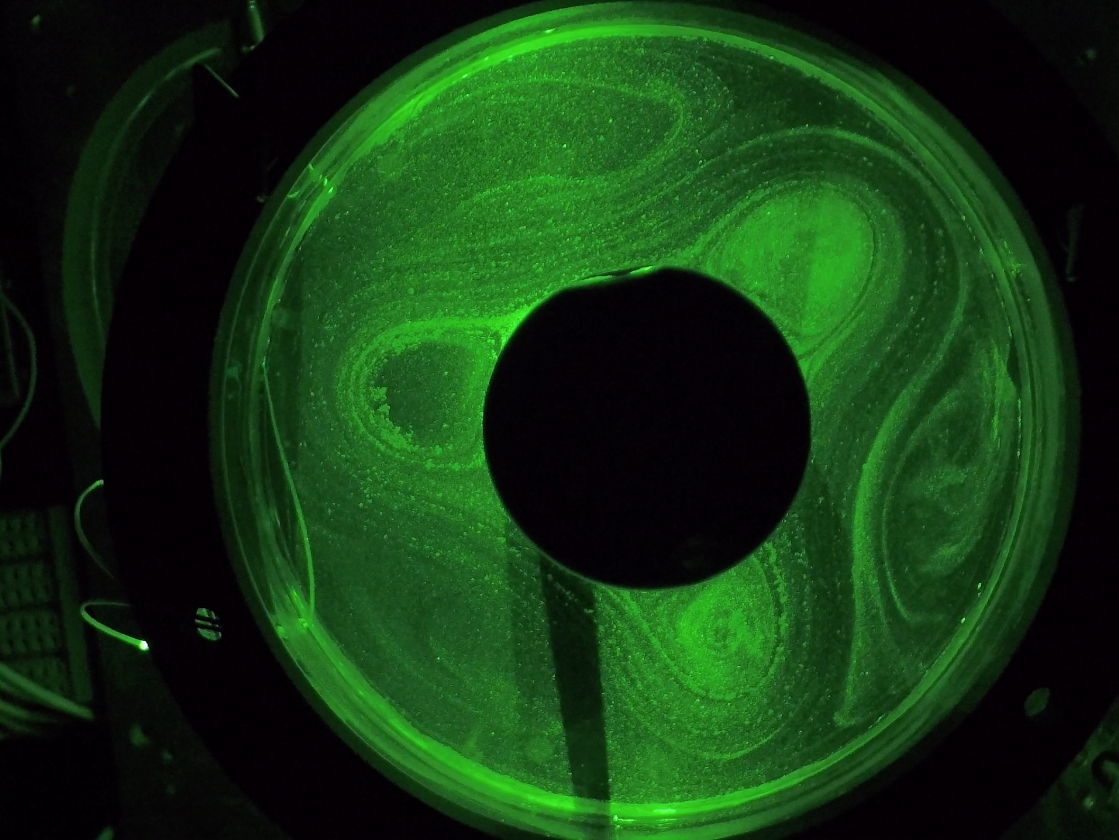} \\
\includegraphics[scale=.18]{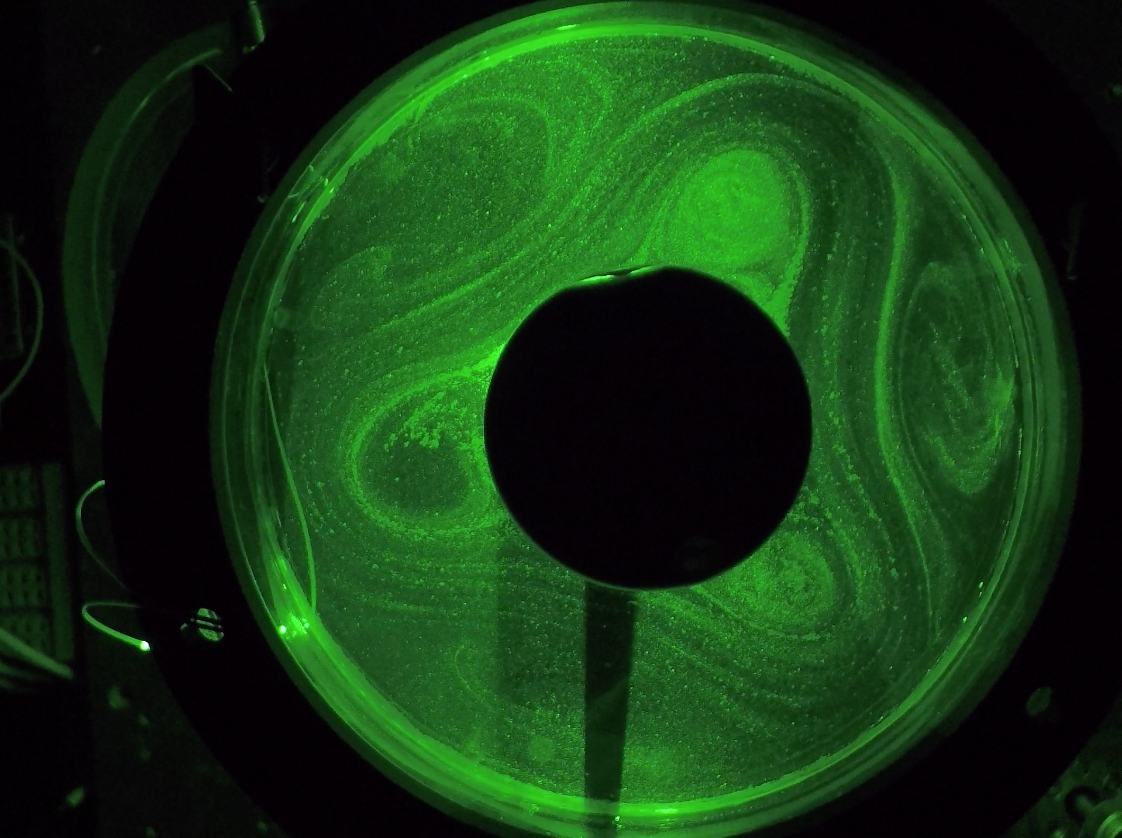} 
\includegraphics[scale=.18]{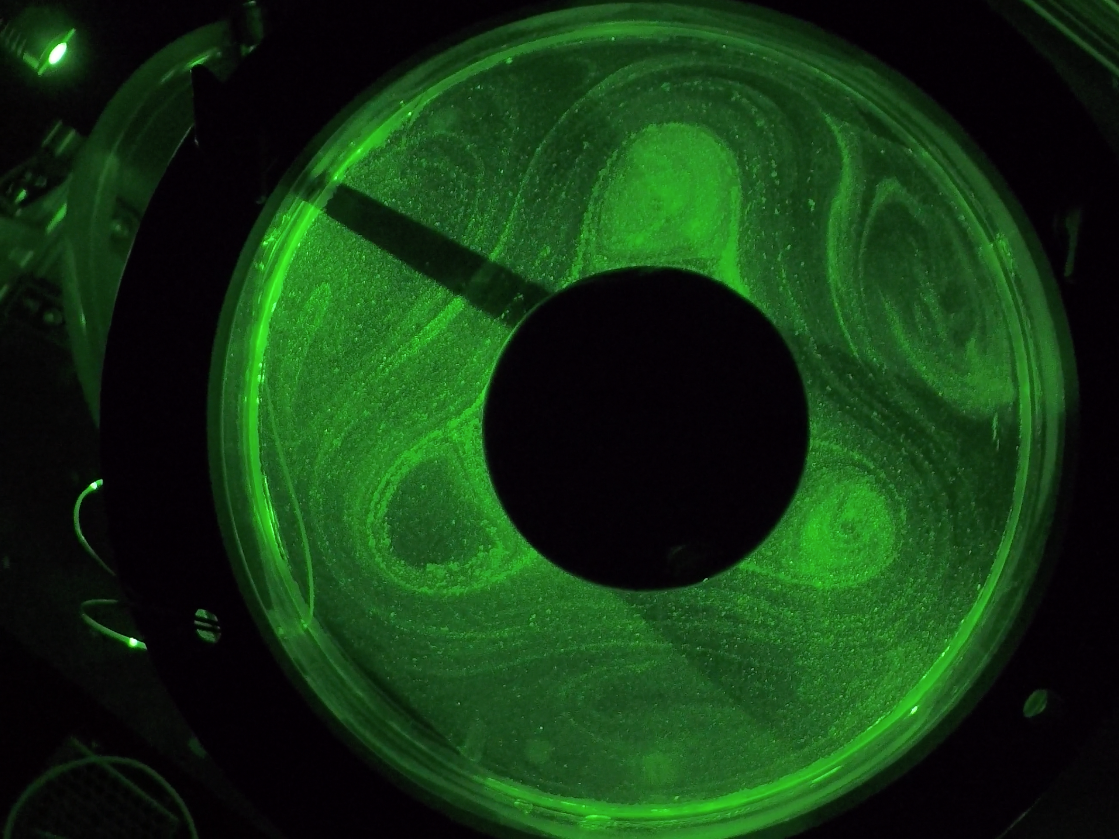}
\caption{Uranine visualization of the baroclinic wave at depth 1 cm from the surface. For this visualization experiment the tank revolution was $5$ rpm, and the temperature difference was $\Delta T=4K$. Pictures have been taken every second revolution. It can be seen that within these 6 revolutions, the baroclinic wave does not change much. The baroclinic jet and the regions within the cyclonic vortices (the areas enclosed by the inner cylinder and the jet) and the anti-cyclonic vortices (the areas enclosed by the outer cylinder and the jet) are also visible. Uranine dye is trapped in these two regions. To avoid a broad shadow behind the inner cylinder, two lasers (Medialas, $75$ and $100$ mW) have been installed in opposite directions. The narrow shadow, visible in 3 images, results from 3 supports mounted in the outer heating ring. Note that PIV particles used in previous measurements can also be seen.}\label{kalli2}
\end{figure*}

\begin{figure}[htbp]
\includegraphics[width=7.5cm,height=7.5cm]{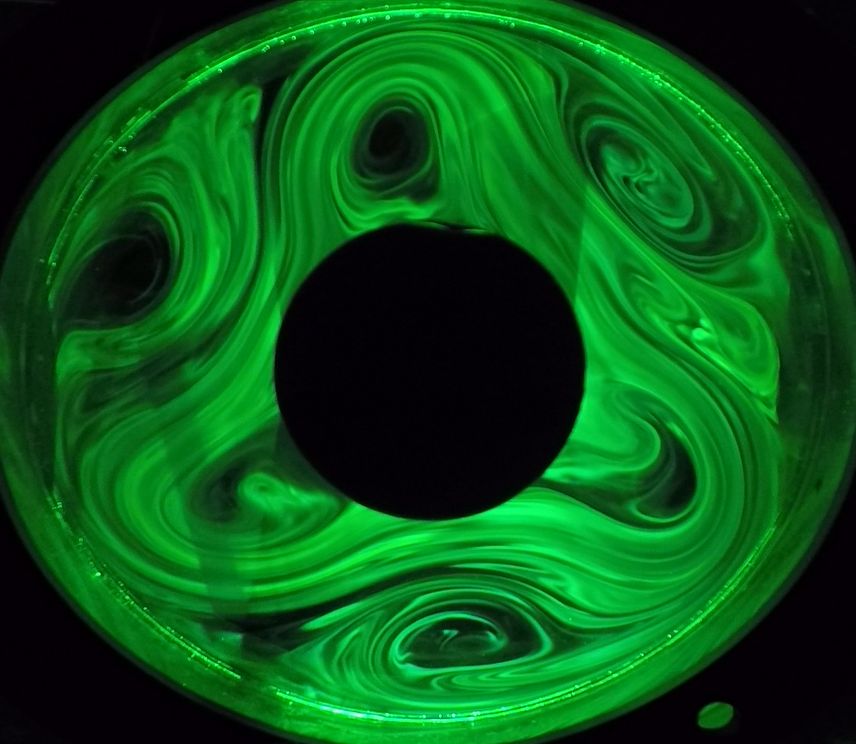} 
\caption{Visualization of the baroclinic wave under the same conditions as in Figure \ref{kalli2}, but with a higher Uranine concentration.}\label{kalli}
\end{figure}

The experiment can be controlled via the rotation rate of the annulus, $\Omega$, and the temperature difference between the cooled inner cylinder and the heated outer annulus, $\Delta T=T_{out}-T_{in}$. For the experiment, the most relevant nondimensional parameters are the Taylor number $Ta$, the thermal Rossby number $Ro$, and the Prandtl number $Pr$. In fact, these numbers are the similarity parameters that connect the experiment with real atmospheric flows. For a qualitative similarity with atmospheric flows, $Ro$ needs to be smaller than one and $Ta$ needs to be very large. For our experiment, we have:
\begin{eqnarray} \label{nondim}
    Ta &=& \dfrac{4 \Omega^2 (b-a)^5}{\nu^2 d} \approx 2.7 \times 10^7 \,, \nonumber\\[.1cm]
	Ro &=& \dfrac{g d \alpha \Delta T}{\Omega^{2} (b-a)^2} \approx 0.1 \,, \nonumber\\[.1cm]
	 Pr &=& \dfrac{\nu}{\kappa} \approx 7 \,,
\end{eqnarray}
where $a,b,d$ are inner and outer radius, and fluid height, respectively, $g=9.81 {\rm m s^{-2}}$ is the constant of gravity, $\nu=1.004\cdot 10^{-6} {\rm m^2 s^{-1}}$ is the kinematic viscosity, $\alpha=0.207 \cdot 10^{-3} {\rm K^{-1}}$ is the thermal expansion coefficient, and $\kappa=0.1434 \cdot 10^{-6} {\rm m^2 s^{-1}}$ is
the thermal diffusivity. The reference temperature for which the values of the fluid properties have been chosen is $20^{\circ} {\rm C}$ (see Table \ref{tab:my_label}).

A Particle Image Velocimetry (PIV) system (from the company Dantec Dynamics) is used to measure the horizontal velocity components $10$ mm below the fluid's surface. Simultaneously, the surface temperature is measured by an infrared camera (InfraTec VarioCAM hr, spatial resolution $640 \times 480$ pixels, thermal resolution $<0.08$K, spectral range $7.5-14 {\rm \mu m}$). The PIV and infrared cameras have been mounted above the annulus and co-rotate with it. To homogenize the temperature and velocity data we applied a linear interpolation by using the MATLAB function \verb!griddata!.

In total, 542 measurements have been taken with a sampling frequency of $0.16$ Hz (2 measurements per revolution). Before taking measurements, the experiments run for $2$ h to guarantee that the flow settled down to a state of equilibrium. Note finally that $Ta$ and $Ro$ have been chosen such that the flow is in a stable wave state with azimuthal wave number $m=3$. \nico{It should be noted that the data sampling rate, commensurable with the rotation frequency, does not lead to some kind of ``stroboscopic'' effects leading to spurious results. In the presented baroclinic wave experiment the velocities are small. The wave drift, as computed later in the kinematic model section, is three orders of magnitude smaller than the angular velocity of the tank rotation. Hence, stroboscopic effects do not play a role in the studied slow quasigeostrophic flow regimes.} 

For a qualitative flow visualization we use Uranine, a fluorescent dye that develops its fluorescent effect and appears green under day or laser light. Figure \ref{kalli2} shows some snapshots for an experiment with parameters $\Omega=5$ rpm and $\Delta T=4$ K. The dye was dripped into the rotating flow. The transport of the dye can then be visualized for a few minutes. After that, the dye distribution becomes more and more homogeneous and the dye filaments disappear. Figure \ref{kalli} shows a visualization of the baroclinic wave under the same conditions as in Figure \ref{kalli2}, but with a higher Uranine concentration



\section{Models}
\label{sec: models}
\subsection{The Empirical Orthogonal Function Analysis}
\label{eof}
In fluid mechanics, analyzing experimentally or numerically obtained spatiotemporal data to extract large-scale structures  is a long-established practice \cite{haken, uhl, holmes}. Fluid evolution can be very complex, however, it has been observed that dominant patterns may exist in the flow field. Several methods exist to this end \cite{uhl,manchotesis,silvina,row1,row2,mezic1,mezic2,kut}, although in this article we will use  the Karhunen-Lo\`{e}ve (K-L) decomposition, also referred to as Principal Component Analysis (PCA) or Proper Orthogonal Decomposition (POD). In the context of atmospheric sciences, this procedure has been referred to as Empirical Orthogonal Function (EOF) analysis. These methods project the spatiotemporal patterns into a basis of orthogonal  functions. These methods are based on a least-square fit, seeking orthonormal modes such that an error function $R_m$ is minimized. More specifically, let us consider that a time-dependent field $q({\bf x}, t)$ is evaluated or measured on a discrete set of points ${\bf x}_i$ (the number of points $i=1,\ldots,n$ may be very large) forming a vector that we refer to as $ q({\bf x}_i,t)=\mathbf{q}(t)  = (q_1(t),\ldots,q_n(t)) \in \mathbb{R}^n$. In our specific case, the points ${\bf x}_i$ correspond to locations within a horizontal plane in the cylinder where velocities or temperatures are measured.  The measured signal $\mathbf{q}$ depends on time, i. e. $q_i(t)$, and although it lives in a space $\mathbb{R}^n$, its motion is constrained to a subspace $\mathbb{R}^m$, where $\mathbb{R}^m \subset \mathbb{R}^n$. The EOF aims to identify the relevant subspace $\mathbb{R}^m$ where the signal evolves. The method is based on a least-square fit of an error function $R_m$, which leads to an eigenvalue problem:
\begin{equation}
C\mathbf{w}_k = \lambda_k \mathbf{w}_k \label{eig}
\end{equation}
where $C$ is a correlation matrix with elements:
\begin{equation}
c_{ij}=\langle q_i,q_j\rangle = \int_{t_0}^{t_0+T}q_i(t) \, q_j(t) \, dt. \label{el_corr}
\end{equation}
The numerical implementation of expression \eqref{el_corr} evaluates the integral using summation. The eigenvalues $\lambda_k$ of expression \eqref{eig} characterize the significance of the
eigenvector $\mathbf{w}_k$ in such a way that the error function $R_m$ becomes:
\begin{equation}
R_m = \sum_{k=1}^m \lambda_k
\end{equation}
The matrix $C$ is symmetric with real eigenvalues and orthonormal eigenvectors. Hence we can project the signal $q_i(t)$ onto the space  spanned by the most significant orthogonal and normalized eigenvectors. Consequently, we get the following reconstruction:
\begin{equation}
   \mathbf{q}(t) = \sum_{k=1}^{m} a_k(t) \mathbf{w}_k\label{model_signal}
\end{equation}
where the time function $a_k(t)$ satisfies:
\begin{equation}
   a_k(t) = \langle \mathbf{w}_k,\mathbf{q}(t) \rangle. \label{amplitude}
\end{equation}
Given the distorted nature of the signal $\mathbf{q}(t)$ produced by measurement artifacts, the amplitudes $a_k(t)$ may be  noisy. In order to  model the experimental signal $\mathbf{q}(t)$ according to equation \eqref{model_signal} we implement a smoothing filter for the amplitudes with the moving averaging method. The EOF method is robust  to noise and finite precision of measurements, due to the application of a least-square-fit potential. For this reason, it also has been implemented in stochastic processes. This method has been used not only in fluid mechanics but also to retrieve global navigation satellite system (GNSS) signals from interference \cite{Dovis,Musumeci,Sharifi}, to suppress surface waves \cite{Liu}, for seismic signals \cite{Serd}, for uncertainty quantification and active learning in partial differential equation models \cite{Rama}, and also it has been used to simulate atmospheric turbulence \cite{Khonina}.
In our work, we will use this analysis for every velocity component and the temperature, as detailed next.

\begin{figure*}[htbp]
	\begin{center}
		a)\includegraphics[scale=.4185]{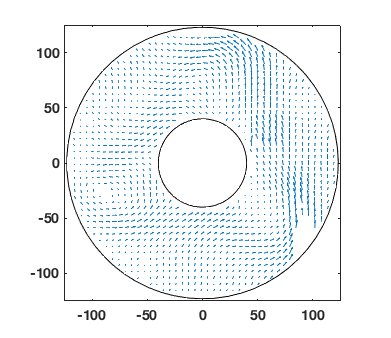}
		e)\includegraphics[scale=.4185]{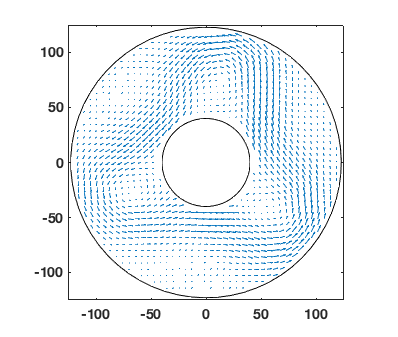}\\
		b)\includegraphics[scale=0.4185]{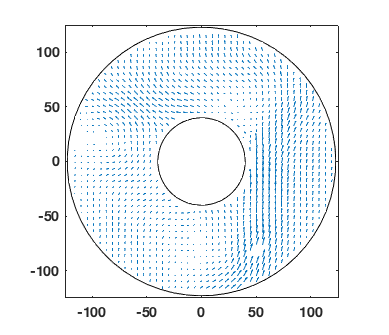}
		f)\includegraphics[scale=0.4185]{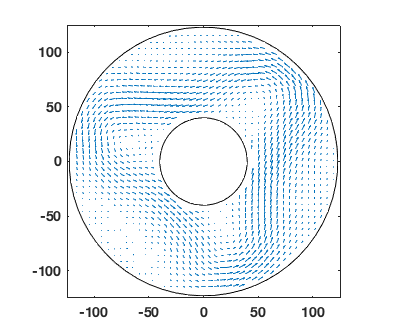}\\
  		c)\includegraphics[scale=0.4185]{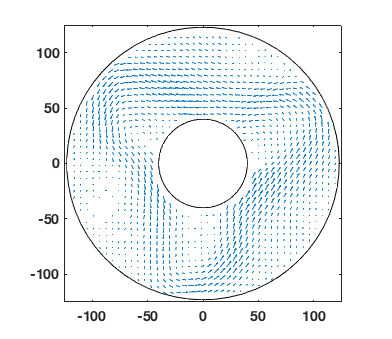}
	  		g)\includegraphics[scale=0.4185]{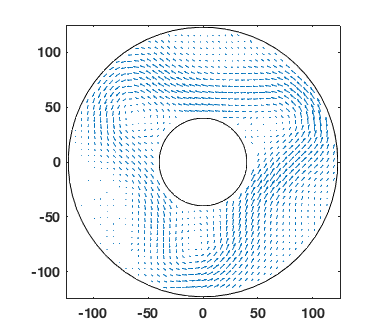}\\
 d)\includegraphics[scale=0.4185]{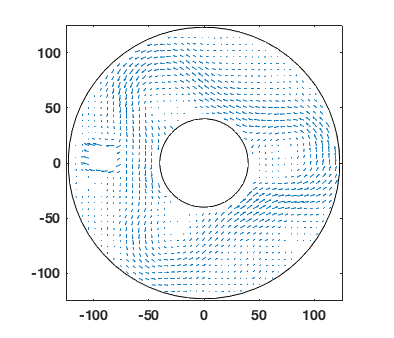}
	 h)\includegraphics[scale=0.4185]{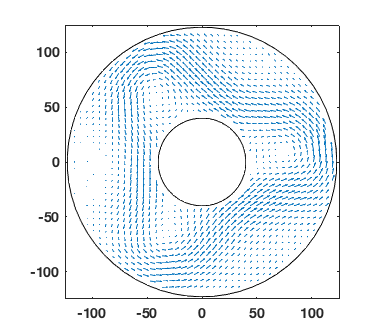}
	\end{center}
	\caption{Representation of experimentally obtained velocities using a PIV system (panels a to d). The snapshots represented correspond to frames 4, 22, 44, and 326 of the experiment. Various gaps and distortions are observed in these snapshots. Representation of filtered velocities using EOF analysis (panels e to h) in the same frames. It is evident that gaps and distortions have been corrected. }\label{veldataeig}
\end{figure*}

\subsubsection{Velocity model}
\label{velocity model}
The experimental setup provides velocity time series data. This data set, due to the measurement procedure, is slightly noisy and has gaps. The first column of Figure \ref{veldataeig} (panels a) to d)) displays a set of velocity fields obtained from the experimental data, which require reconstruction.   The second column (panels e) to h))  shows the reconstruction obtained with the EOF analysis.  This is done with a decomposition of the field on the eigenfunctions of the system \eqref{eig} with the highest eigenvalues. In this case, we have considered $m=5$. Once these orthogonal functions are computed, the time series  is obtained  following equation \eqref{model_signal} after smoothing the amplitudes $a^l$. \nico{Note that the noisiness of the velocity data is mainly due to poor laser light conditions as a consequence of moving shadows in the observation area. As mentioned in section \ref{sec:experimemtal_setup} it is not due to surface waves. For the experiment discussed the free surface can be considered as flat. Hence the data gaps have no physical origin and have thus no impact on the dynamic of the flow.}

\subsubsection{Temperature model}
\label{temperature model}

\begin{figure*}[htbp]
	\begin{center}
        a)\includegraphics[scale=0.35]{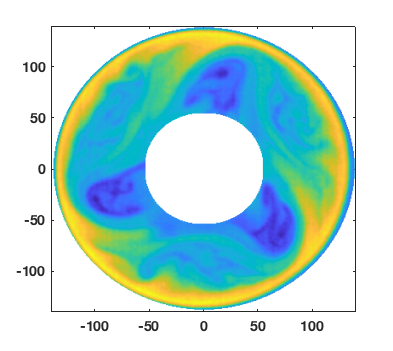}
            b)\includegraphics[scale=.35]{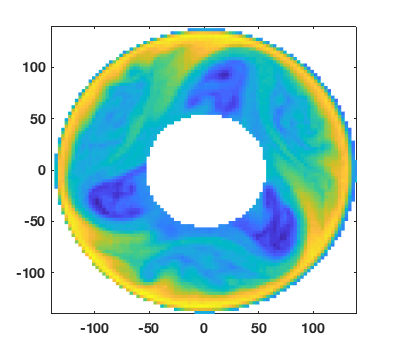}
      c)\includegraphics[scale=.35]{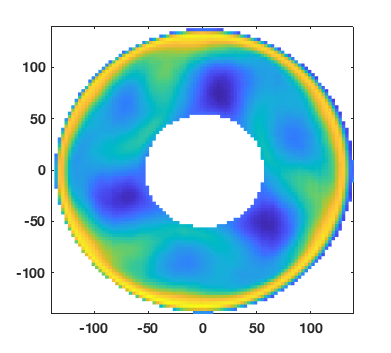}\\
		d)\includegraphics[scale=0.35]{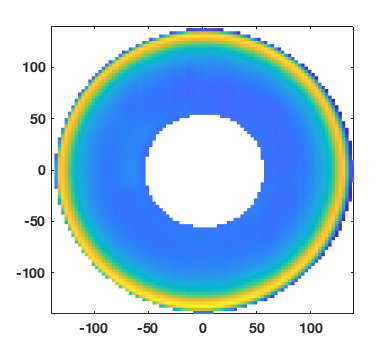}
		e)\includegraphics[scale=.35]{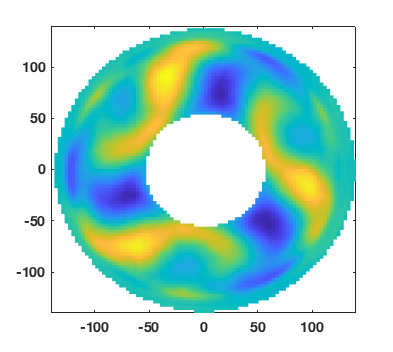}
  	f)\includegraphics[scale=.35]{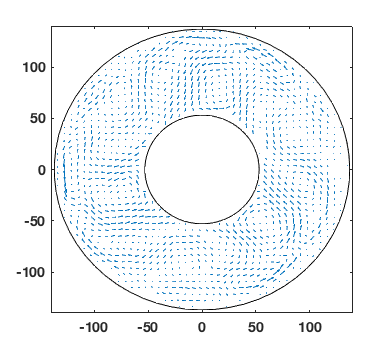}\\
  	g)\includegraphics[scale=.7]{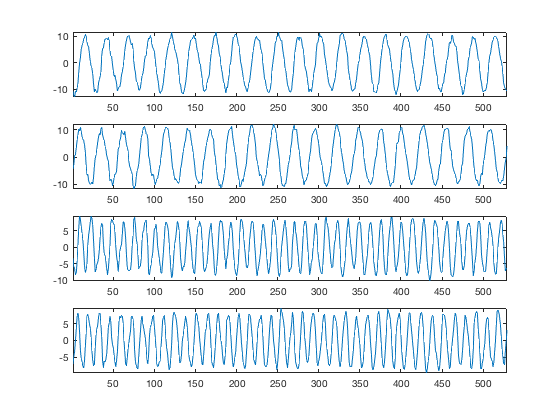}
   	\end{center}
	\caption{a) Representation of the surface temperature field measured in the 5th snapshot; b) A coarser representation of the field in a) produced to perform the EOF analysis; c) filtered version; d) temporal average of the signal in each point; e) EOF analysis applied to the signal obtained after subtracting the time average in d) from the signal in b); f) velocities obtained from the signal in e); g) time amplitudes associated to the four EOF with highest eigenvalues.}\label{veltempeig}
\end{figure*}

The experimental setup provides temperature time series data as illustrated in Fig. \ref{veltempeig}. Panel a) shows an output of the experimentally measured temperature field for the 5th snapshot. The field has a very high resolution, which is diminished in panel b). This is a necessary step to apply the EOF method; otherwise, the correlation matrix in expression \eqref{eig} has an unmanageable dimension. Panel c) shows the filtered version after applying the EOF analysis. In this analysis, the signal denoted by ${\bf q}(t)$ in section \ref{eof} is constructed from the time series of snapshots, such as the one in panel b), after subtracting the temporal average displayed in panel d) at each point. Panel e) shows the resulting pattern after applying the EOF analysis to this signal and retaining the six most significant modes. Panel g) displays the time series of four of the most significant modes. Finally, panel f) presents the obtained velocity field using expressions \eqref{veltermx} and \eqref{veltermy}.  Next, we explain the rationale behind these expressions.

The thermal Rossby number in Eq. \eqref{nondim} contains the so-called \emph{thermal wind balance}. In its standard form, $Ro$ reads
\begin{equation}
    Ro=\frac{U}{f L}.
\end{equation}
where $U$ and $L$ are, respectively, the characteristic velocity and length scales, and $f$ is the Coriolis frequency. When $U$ is replaced by the thermal wind
\begin{equation} \label{UT}
    U_T=\frac{g d}{b-a} \frac{\alpha \Delta T}{f},
\end{equation}
the Rossby number transforms into the form given in Eq. \eqref{nondim}.
The thermal wind relies on the geostrophic and hydrostatic approximation:
\begin{eqnarray}    
    f u &=& -\frac{1}{\rho_0} \frac{\partial p}{\partial y},\\[.1cm]
    f v &=& \frac{1}{\rho_0} \frac{\partial p}{\partial x},\\
    \frac{\partial p}{\partial z} &=& -\rho g.
\end{eqnarray}
Differentiating the geostrophic approximations with respect to $z$ and replacing the pressure by density using the equation of state for fluids, $\rho=\rho_0 (1-\alpha T)$, we find
\begin{eqnarray}    
     \frac{\partial u}{\partial z} &=& -\frac{g \alpha}{f} \frac{\partial T}{\partial y},\\[.1cm]
     \frac{\partial v}{\partial z} &=& \frac{g \alpha}{f} \frac{\partial T}{\partial x}.
\end{eqnarray}

\begin{figure*}[htbp]
a) \includegraphics[scale=.209]{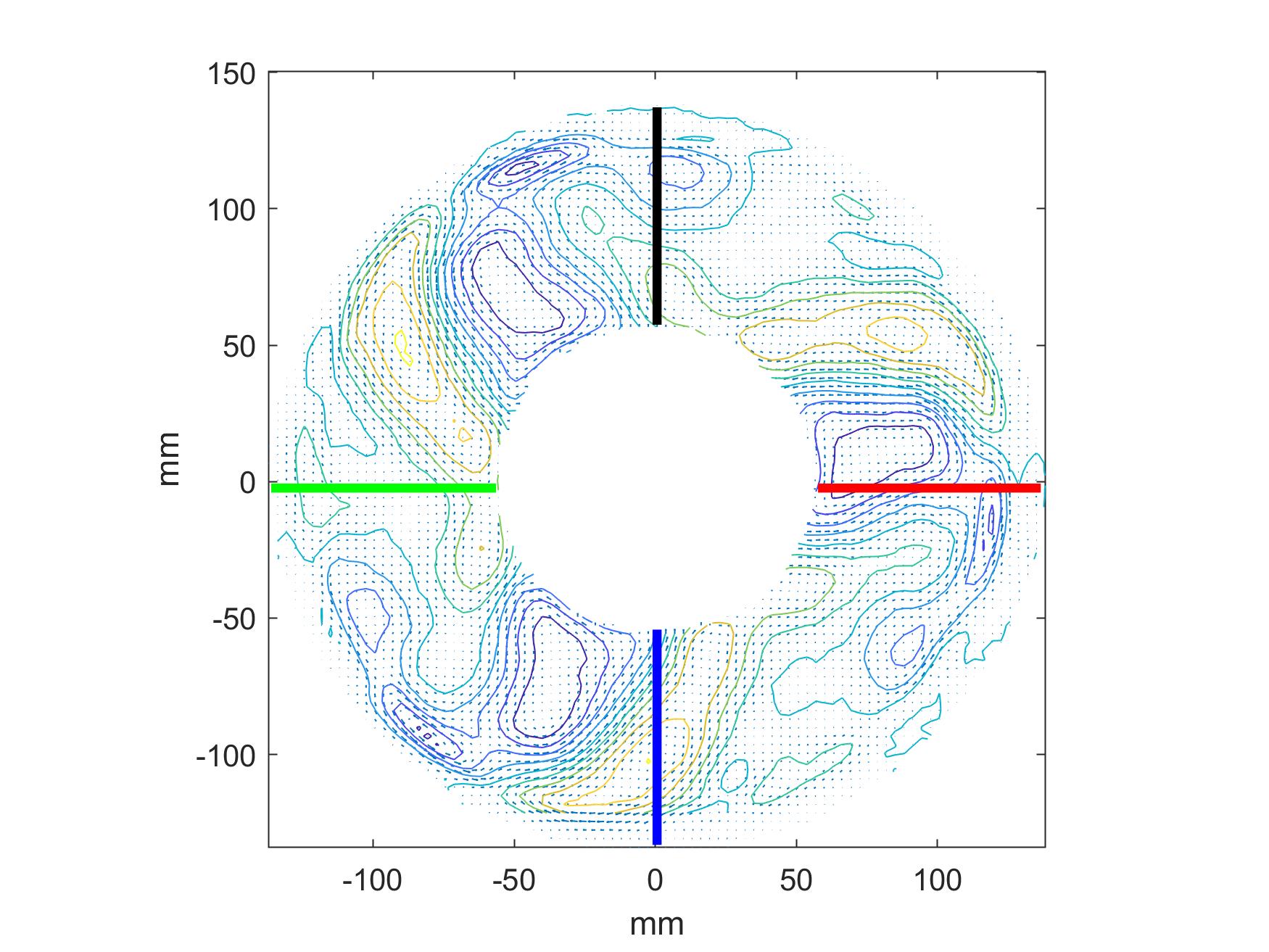} \\
b) \includegraphics[scale=.18]{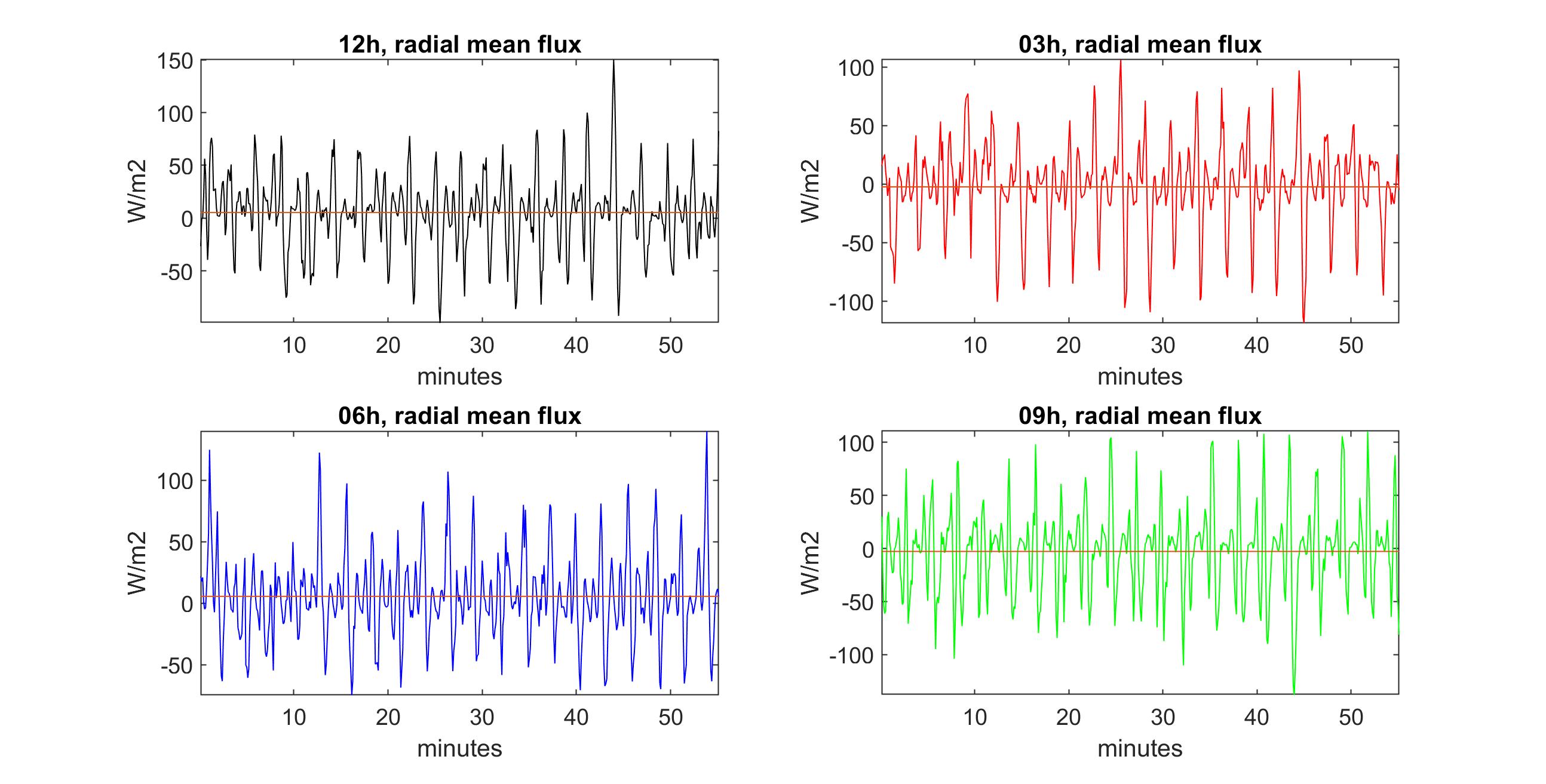} \hspace{.2cm}
\caption{a) Snapshot of the annulus temperature field and the corresponding velocity field constructed via $\eqref{thermal}$. b) Time series of the radial eddy temperature flux $\rho_0 c_p \overline{U_r' T'}$ averaged along the four lines shown in a). The horizontal red line shows the time mean value.}\label{flux}
\end{figure*}

Finally, integrating these two equations with respect to $z$ from the top of the bottom Ekman layer to the free surface, assuming that the Ekman layer flow, as well as the horizontal gradients of the temperature in the Ekman layer, are small, and also that $T$ does not strongly depend on $z$, we find a first estimate of the surface flow from:
\begin{eqnarray}
\label{thermal}
    U_{x} &=& -\frac{g \alpha d}{f} \frac{\Delta_y T}{\Delta y}, \label{veltermx}\\[.1cm]
    U_y &=& \frac{g \alpha d}{f} \frac{\Delta_x T}{\Delta x},\label{veltermy}
\end{eqnarray}
where $U_{x}$ and $U_y$ represent velocity components in Cartesian coordinates, coinciding with $U_{\phi}$ and $U_r$, which denote the azimuthal and radial velocity components, respectively, along the green, blue, red, and black lines marked in Fig. \ref{flux} a).

Note that radial temperature flux connected to the so-constructed geostrophic velocities, see Eq. \eqref{thermal}, is zero in the long term. The time mean of the radial temperature flux can be written as
\begin{equation}
    \overline{(\overline{U_r} + U_r') (\overline{T}+T')}=\overline{\overline{U_r} \overline{T}}+\overline{\overline{U_r} T'}+\overline{U_r' \overline{T}}+\overline{U_r' T'},
\end{equation}
where the bar denotes the time-mean and the prime the eddy component. The first term on the right-hand side vanishes since the mean radial flow is zero. The second and third term is zero since the time mean of the eddies vanishes. In general, in the last term, the radial eddy temperature flux is nonzero. A significant contribution is due to the fact that the eddy temperature field shows a phase shift with respect to the velocity field. The Eady model \cite{Eady49} of baroclinic instability shows this very clearly (see, e.g., figure 10 in \cite{Harlander12}). Per construction, such a phase shift does not exist for Eq. \eqref{thermal} and hence we cannot expect a long-term temperature flux. Nevertheless, locally the temperature flux shows a high temporal and spatial variability due to the pro-grade propagating eddies in the annulus. 

In Fig. \ref{flux}b) we display the time series of the radial temperature flux averaged along the four lines shown in Fig. \ref{flux}a. The mean flux and standard deviation of these four cases are about $1.5$ W/$m^2$ and $36$ W/$m^2$. However, the time and spatial mean over the whole annulus, $\rho_0 c_p \langle \overline{U_r' T'} \rangle $ with $\rho_0 = 10^3$ kg m$^{-3}$ and $cp = 4.187$ J kg$^{-1}$ K$^{-1}$ is just $0.0023$ W/$m^2$. Note that the flux in real quasi-geostrophic flows is, apart from the mentioned phase relation between velocity and temperature, provided by ageostrophic flow components and within boundary layers. Still, the geostrophic flow constructed via (\ref{thermal}) is a reasonable approximation of major flow features. As we will see, this approximation is useful for understanding properties of nonlinear transport and mixing of the flow in the differentially heated rotating annulus.
Finally we should add that \nico{Rodda et al.} \cite{Rodda_etal22} found that
\begin{equation}
    U_T=\frac{g d}{b-a} \left( \frac{\alpha \Delta T}{f} \right)^\zeta,
\end{equation}
with $\zeta=0.55$ gives a somewhat better fit to the experimental surface flow data over a range of $Ro$ and $Ta$ values than $\zeta=1$ used in (\ref{UT}). However, for the kinematic model developed here, the smaller exponent does not play a significant role.


\subsection{The kinematic model}
\label{kinematic_model}

Kinematic models have a long history in the geophysical fluid dynamics community and have provided a simple approach to studying Lagrangian transport and mixing. Inspired by kinematic models proposed in \cite{AGP,Castillo,Samelson,Wiggins,Sommeria1,Mancho2,Sommeria2}, we propose a kinematic model in the form of a 2D streamfunction $\Psi$ with the capability of recovering the characteristic transport features observed in the laboratory experiments. More specifically, the 2D streamfunction $\Psi$ is fitted from the experimental data and used to extend the model in domains where the data are noisy, distortedly measured, or simply missing.



\begin{figure*}[htbp]
\begin{center}
a)\includegraphics[scale=0.4]{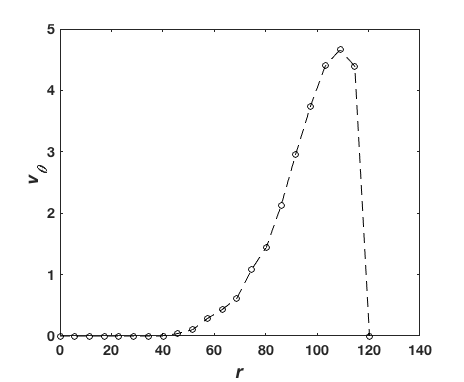}
b)\includegraphics[scale=0.4]{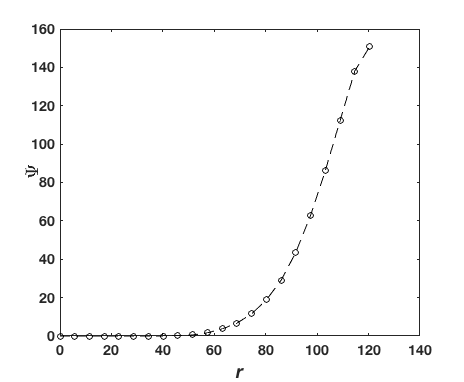}\\
c)\includegraphics[scale=0.4]{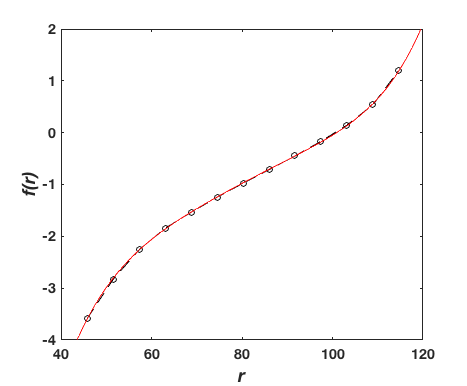}
d)\includegraphics[scale=0.4]{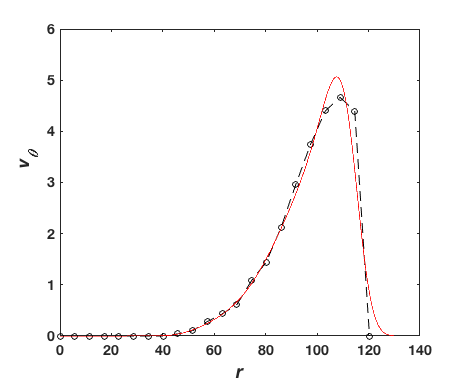}\\
e)\includegraphics[scale=0.4]{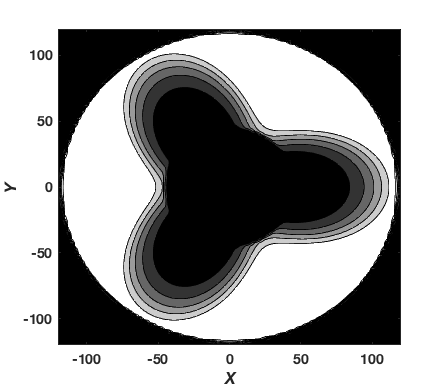}
f)\includegraphics[scale=0.4]{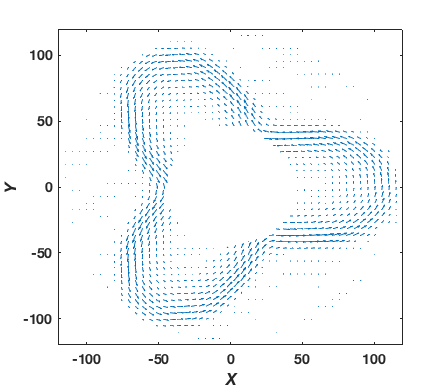}
\end{center}
\caption{a) Experimental measurement of $v_{\theta}$  represented versus the $r$ coordinate at a specific snapshot (frame 222); b) numerical evaluation of the streamfunction using expression \eqref{integral} and experimental values of $v_{\theta}$; c) evaluation of the function $f(r)$ given in Eq. \eqref{f_function} (black circles), together with the fitting of a quintic polynomial (red line); d) Comparison of the velocity data from the experiment (black circles) along the line $\theta=0$, with the velocities obtained from the streamfunction $\Psi(r)$ given in Eq. \eqref{psifunction} (red line); e) representation of the streamfunction with $\theta$ dependency; f) snapshot of velocities at $t=0$ according to expressions \eqref{velfromtstream}.}
\label{datakm}
\end{figure*}

Figure \ref{datakm} summarizes the steps taken to fit the streamfunction, $\Psi$, and therefore velocities of the kinematic model, from velocity measurements. Panel a) represents a snapshot of the $v_y$ velocity component along the axis $y = 0$, with $x \geq 0$, and thus at $\theta = 0$ in polar coordinates. 
This is the velocity obtained from the experimental data at a chosen time that approximately satisfies that, along the selected axis, $v_x=0$. 
In the experiment, velocities are obtained in Cartesian coordinates. However, at this specific angle, $\theta = 0$, we observe that $v_x=v_r$ and $v_{y} = v_{\theta}$; thus, we consider that $v_{y}$ is an estimate of the angular velocity, $v_{\theta}$.
Units  are respectively millimeters per second for the velocity and millimeters for the $r$ coordinate.

A streamfunction will be proposed that depends only on the radius, and in a second step we will make this streamfunction to depend on the angle.  In order to find the dependency of $\Psi$ with respect to
$r$, we solve the equation:
\begin{equation}
v_{y}(x>0, y=0) = -\dfrac{\partial \Psi}{\partial x}      \nonumber
\end{equation}
or equivalently, assuming that $\Psi$ depends only on $r$
\begin{equation}
v_{\theta}(r) = -\dfrac{\partial \Psi}{\partial r}
\end{equation}
Therefore, the streamfunction is given by:
\begin{equation}
\Psi(r) = -\int_{0}^{r} v_{\theta}(r^\prime) \, dr^\prime 
\label{integral}
\end{equation}
To obtain $\Psi$ from Eq. \eqref{integral}, the velocity data in panel a) is numerically integrated using the trapezoidal method. Panel b) displays the resulting numerical function evaluated at discrete points, exhibiting similarity to a hyperbolic tangent, as seen in the model used in \cite{Samelson}. Indeed, we search for an explicit expression   for the streamfunction $\Psi$ assuming   it follows a hyperbolic tangent as follows:
\begin{equation}
\Psi(r) = A\left(\dfrac{\tanh(f(r))+1}{2}\right)
\end{equation}
To propose this function, we have transformed the hyperbolic tangent, which takes values between $[-1,1]$, for it to take values between $[0,A]$. Here $A$ is the maximum value of the streamfunction, and $f(r)$ is a function that needs to be adjusted. In particular, we consider:
\begin{equation}
f(r) = \textrm{arctanh}\left(\dfrac{2\Psi(r)}{A}-1\right) \;.
\end{equation} 
In panel c)  this function is represented \nico{ using $A = 150.67$ and the experimental data   for $ \Psi(r)$ (black circles). The red line represents the adjustment to the data with a non-linear least squares method. In particular we use  a quintic polynomial fit of the form}:
\begin{equation}
f(r) = a_5 \, r^{5} + a_4 \, r^{4} + a_{3} \, r^{3} + a_{2} \, r^{2} + a_{1} \, r + a_{0} \;,
\label{f_function}
\end{equation}
\nico {The adjustment gives} $a_5=1.258 \cdot 10^{-8},\ a_4=- 5.068\cdot 10^{-6},\ a_{3} = 8.212 \cdot 10^{-4},\ a_{2}=-0.06706,\ a_{1} =  2.812,\ a_{0}=-50.86$. \nico{The  root-mean-square deviation obtained from the experimental data and the fitting is 0.0446}. This gives us a \nico{model} function $\Psi$ that depends only on $r$. 
\begin{equation}
\Psi (r) = \dfrac{A}{2}\left(\tanh (a_{5}  r^{5} + a_{4}  r^{4} + a_{3}  r^{3} + a_{2}  r^{2 }+ a_{1}  r + a_{0}) + 1\right) 
\label{psifunction}
\end{equation}
Panel d) confirms that along the line $\theta=0$, there is a good agreement between the velocity data \nico{(black circles)} and the velocities obtained from the proposed analytical streamfunction \nico{(red line)}. 

\begin{figure*}[htbp]
\begin{center}
a)\includegraphics[scale=0.3]{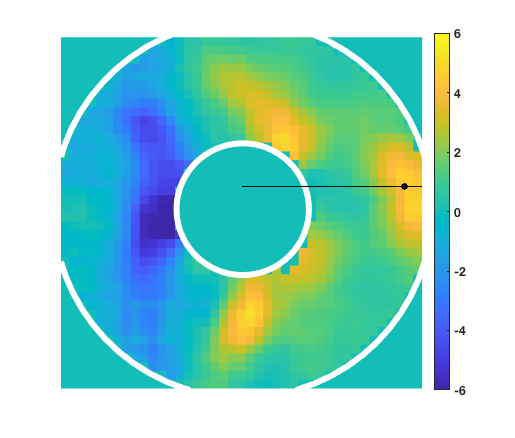}
b)\includegraphics[scale=0.3]{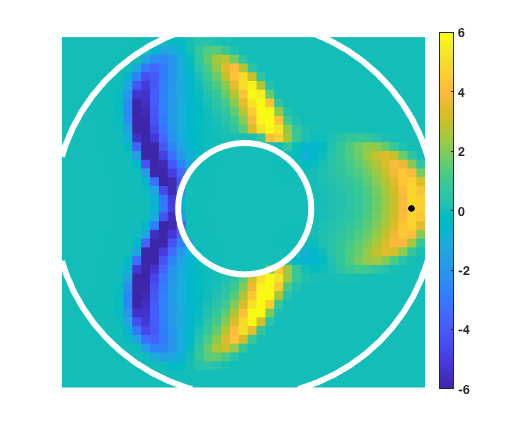}\\
c)\includegraphics[scale=0.3]{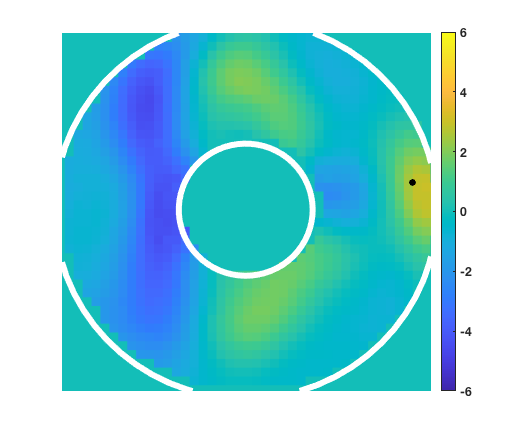}
d)\includegraphics[scale=0.3]{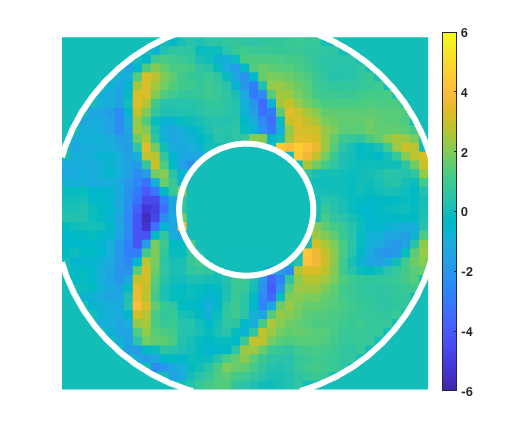}\\
e)\includegraphics[scale=0.5]{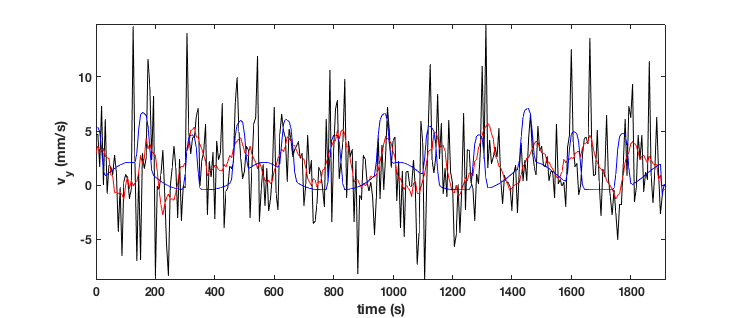}
\end{center}
\caption{\nico{a) Experimental measurement of $v_{y}$  represented  at a specific snapshot (frame 222). The black line highlights the location for the graph displayed in figure \ref{datakm}a); b)  representation at $t=0$ of $v_y$ adjusted with the kinematic model; c) representation at frame 222 of the filtered $v_{y}$ using EOF analysis; d)  difference between  the experimental $v_{y}$ and the one obtained from the kinematic model; e) $v_{y}$ time series at positions marked with black dots for the experiment (black), kinematic model (blue), filtered EOF analysis (red).}}
\label{comp}
\end{figure*}

The streamfunction in Eq. \eqref{psifunction} depends only on $r$, however, the streamfunction leading to velocities fields like those in Figure \ref{veldataeig} depends also on the angle $\theta$. We propose a dependency on  the angle in such a way that for different $\theta$-directions the maximum gradient of the streamfunction in Eq. \eqref{psifunction}, i.e. where the jet is placed,  becomes shifted between the outer and the inner cylinder following the 3-mode oscillation. 
This is accomplished by substituting $r$ in Eq. \ref{psifunction} with $r+\alpha$, where $\alpha$ is associated with the shift occurring between the cylinders. For example, in panel e), the streamfunction is represented, showing that the location of the highest gradients of the streamfunction (the jet) undergoes an oscillatory shift towards both the inner and outer cylinders. This shift is achieved by introducing a dependency of the displacement $\alpha$ on the angle $\theta$.
Therefore, we achieve a model streamfunction $\Psi (r, \theta)$ by replacing $f(r)$ with $f(r+\alpha(\theta))$, where:\nico{
\begin{equation}
 \alpha (\theta) = \alpha = c_{1}\cos(3\theta)+c_{2},\quad c_{1} = -30, \quad c_{2} = 30
\end{equation}
}
These choices adjust perfectly to the setting. 
Given that, in the experiment, as illustrated in panels a) to d) of Figure \ref{veldataeig}, the entire velocity pattern is rotating, expression \eqref{psifunction} requires further adjustments. We achieve this by shifting $\alpha$, at a frequency $\omega$. Based on the experimental time series dataset, this frequency is adjusted to $\omega = -0.039 \, s^{-1}$.
We therefore propose a final expression for $\alpha$ depending also on $t$, as follows:
\begin{equation}
 \alpha (\theta) = \alpha = c_{1}\cos(3\theta +\omega t) + c_{2} \,,
\label{alfa} \end{equation}

Now we consider an additive perturbation $\beta$ for the streamfunction, whose role is detailed in the next subsection. It is as follows:
\begin{eqnarray}
\Psi &=& \dfrac{A}{2} \left( \tanh \left( a_{3}(r+\alpha)^{3}+a_{2}(r + \alpha)^{2}+a_{1}(r+\alpha)+a_{0} \right) +1 \right) \nonumber \\&+& \beta(t) \, r \, h(r)
\label{psiper}
\end{eqnarray}
where
\begin{equation}
\beta(t) =  c_3 \sin\left(\gamma t+\frac{\pi}{4}\right)^4-c_4
\label{beta0}
\end{equation}
$\gamma=0.0069$ and $h(r)$ is a step-like function that kills the perturbation for $r$ within the inner cylinder and $r$ above the outer cylinder. The explicit form of this function is:
\nico{
\begin{equation}
h(r) = \left(\dfrac{\tanh (r-45)+1}{2}\right)\left(\dfrac{-\tanh(r-118)+1}{2}\right)	
\end{equation}
}
The perturbing term is heuristically adjusted to produce perturbations that align with the experimental data. A discussion on  specific selections for $c_3$ and $c_4$ is presented in Section IV. 

Panel d) shows a snapshot of the velocities obtained from the streamfunction   \eqref{psiper}. These are obtained after considering  the streamfunction   in Cartesian coordinates using $r = \sqrt{x^{2}+y^{2}}$ and $\theta = \arctan (y/x)$, which  yields:
\begin{eqnarray}
	 v_{x} &=& -\dfrac{\partial \Psi}{\partial y} = -\left(\dfrac{\partial \Psi}{\partial r} \dfrac{\partial r}{\partial y} + \dfrac{\partial \Psi}{\partial \theta} \dfrac{\partial \theta}{\partial y}\right) \label{velfromtstream}\\[.1cm]
	v_{y} &=& \dfrac{\partial \Psi}{\partial x} = \left(\dfrac{\partial \Psi}{\partial r} \dfrac{\partial r}{\partial x} + \dfrac{\partial \Psi}{\partial \theta} \dfrac{\partial \theta}{\partial x}\right)\nonumber
\end{eqnarray}

\nico{Figure \ref{comp} completes this section by comparing the outputs in space and time of the kinematic model with the velocity field obtained from the experiment. Panel a) shows the spatial representation of the measured $v_y$ with the PIV system at time frame 222. The black line highlights the location of the measurement used for adjusting the kinematic model. The colorbar, which is the same from panels a) to d), ranges between -6 and 6 mm/s. The black dot marks the position associated with the time series shown in panel e). Panel b) displays the velocity $v_y$ obtained from the kinematic model at time $t=0$. This time in our model is set to coincide with the experimental frame 222. 
Again, the black dot marks the location associated with the time series. Panel c) displays, at frame 222, the velocity $v_y$ obtained after applying the EOF analysis described before to the full experimental dataset. It is noticed that filtering regularizes the pattern but also weakens it. Panel d) displays the difference between the experimental field and the one obtained from the kinematic model at the same time. A very good agreement, {\em i. e.} small error, is observed along the semi-axis used to fit the kinematic model ($y=0$ and $x>0$). The $L^2$ norm of this error function in the whole domain is 51.6 mm/s. Finally, panel e) displays the $v_y$ time series at positions marked with the black dots linked to the raw experimental data (black), the filtered data (red), and the kinematic model (blue).}

\subsection{The transport model}

The velocity models form the basis of the transport models. Indeed, in the purely advective approach, passive  scalars follow trajectories, denoted as ${\bf x}(t) = (x(t), y(t))$, which are  the solutions to the system:
\begin{equation}
\frac{d{\bf x}}{dt}={\bf v}({\bf x},t),  \label{sd}
\end{equation}
where ${\bf v}=(v_x,v_y)$ are the velocities obtained according to different assumptions in subsections III A and III B. 

In subsection \ref{eof} velocities are provided as datasets and therefore the right-hand side of the system in \eqref{sd} requires interpolation. We use cubic interpolation which is discussed to be an appropriate choice (see \cite{interp}). 

Regarding the kinematic model in subsection \ref{kinematic_model}, the  additive perturbation for the streamfunction distorts the velocity field in such a way that solutions to the system \eqref{sd} that typically are closed trajectories forming what in dynamical system theory is called invariant tori,  become more erratic and chaotic due to the presence of resonant tori \cite{kam1,kam2,kam3,nek,Mancho2}. The condition for resonant or non-resonant tori  relates  the frequencies of the closed trajectories for the unperturbed  stationary system  with the frequency of the perturbation given in expression  \eqref{beta0}.
A condition for the destruction of a closed trajectory is that the quotient of its period with that of the perturbation is close
 to a rational number \cite{velasco, Mancho2}. In any case, the perturbation includes a quartic exponent, which couples several fundamental perturbing frequencies.

\section{Results and Discussion}
 \label{sec: results} 


This section compares the performance of the constructed velocities using different approaches to describe the experiment. The primary focus is on examining their ability to reproduce patterns, such as those observed in Figure \ref{kalli2}. We assume that the Uranine dye used for visualization in this figure closely follows fluid parcel trajectories, denoted as ${\bf x}(t)$, which in the purely advective approach  represent the solutions to the system \eqref{sd}. The patterns are made visible in Figure \ref{kalli2} 
through the alignment of the Uranine dye with certain mathematical structures that are present in dynamical systems such as that in equation \eqref{sd}. 

These structures correspond to attracting material curves, which are associated with the unstable manifolds of hyperbolic trajectories. Hyperbolic trajectories exhibit highly contracting and expanding directions related to stable and unstable manifolds, respectively. A cluster of a passively advected scalar  initially placed in the neighborhood of a hyperbolic trajectory evolves over time, contracting along the stable direction and expanding along the unstable direction. After a sufficiently long period, this cluster aligns with a complex curve known as the unstable manifold, which  for this reason is referred to as an attracting material curve.  
Mathematically, this behavior is justified by the Inclination Lemma, which roughly states that the forward-time evolution of a cluster intersecting the stable manifold of a hyperbolic trajectory will eventually get aligned with the unstable manifold. 
Hyperbolic trajectories, along with their stable and unstable manifolds, are crucial geometric elements for describing transport evolution in phase space, which coincides with the physical space in advection. Additionally, there are regions in the flow where dispersion does not occur. These regions are associated with fluid parcels that remain confined, forming closed trajectories known as tori. While tori find realistic implementations in ocean eddies, it is important to note that they are mathematically idealized objects and may not be present in aperiodic and non-regular fluid flows.   In the context of geophysical flows, the reported structures are collectively referred to as Lagrangian Coherent Structures.

In time-dependent dynamical systems such as that in Eq. \eqref{sd} hyperbolic trajectories, their stable and unstable manifolds, and tori may be visualized with the method of Lagrangian Descriptors (LDs). 
This is a trajectory-based scalar diagnostic introduced in [52], serving as a fundamental component for defining the concept of a distinguished trajectory. Later, \cite{Mendoza} showed that LDs could play a major role in the description of geophysical flows, and \cite{cnsns} provided several generalizations of the method. Formal results about Lagrangian Descriptors are found in \cite{carlos,lopesino2017,gg2018b}. Since then, LDs have been used in many applications in geophysical flows \cite{alvaro2,delacamara2012,Curbelo,ggrmcw16,ramos2018,garciasanchez2020} and chemistry \cite{Craven, Agaoglou1,Garcia2, Agaoglou2,Agaoglou3}. 

\begin{figure}[htbp]
\begin{center}
a) \includegraphics[scale=0.45]{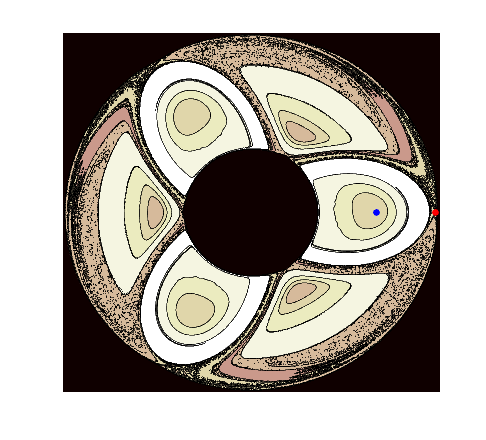}\\
b) \includegraphics[scale=0.45]{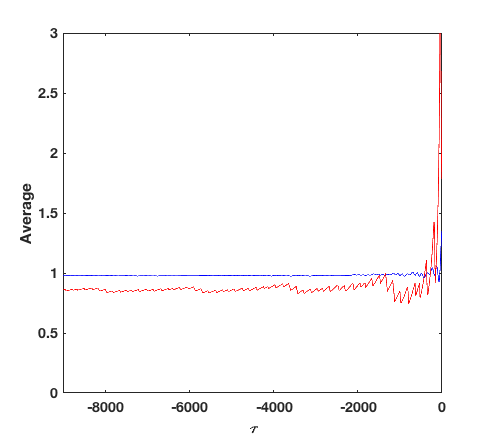}
\caption{a)   
Attracting material curves are highlighted using contour lines of $|| M^{(b)}||$ computed at $t_0 = 0$ s and $\tau = 6600$ s, based on the kinematic model with $c_3=0.4$ and $c_4=0.2$. The blue dot represents an initial condition in a domain with confined behavior (tori), while the red dot represents an initial condition in a domain with chaotic behavior; b) evolution of the averages of $M^{(b)}/\tau$ versus $\tau$ for the selected initial conditions is shown. The red line, associated with the red dot, does not converge within the displayed $\tau$ interval, while the blue one does. The former corresponds to a chaotic trajectory covering the entire domain, while the latter represents a confined trajectory on a torus. }
\label{resultsb}
\end{center}
\end{figure}

The particular LD that we use in this work is a function referred to as $M$ \cite{Madrid,Mendoza,cnsns}, and is defined as follows:
\begin{equation}
    M({\bf x}_0,t_0,\tau) = \int_{t_0-\tau}^{t_0+\tau} \|{\bf v}({\bf x}(t),t)\|\ dt
    \,,
\label{M}
\end{equation}
where $||\cdot||$ stands for the modulus of the velocity vector and $\tau$ is the integration time. At a given initial time $t_{0}$, the function $ M({\bf x}_0,t_0,\tau)$ measures the arclength traced by the trajectory starting at $\mathbf{x}_{0} = \mathbf{x}(t_{0})$ as it evolves forward and backward in time for the time interval $\tau$.  It has been demonstrated in the literature \cite{cnsns,lopesino2017} that the stable and unstable manifolds of hyperbolic trajectories in \eqref{sd} align with singular features observed in the field represented by \eqref{M} over a sufficiently long integration period $\tau$.
The integral in Eq. \eqref{M} can be split into two terms, $M = M^{(f)} + M^{(b)}$, where:
\begin{eqnarray}
	M^{(f)}({\bf x}_0,t_0,\tau) &=& \int_{t_0}^{t_0+\tau} \|{\bf v}({\bf x}(t),t)\|\ dt,\\[.1cm]
  M^{(b)}({\bf x}_0,t_0,\tau) &=& \int_{t_0-\tau}^{t_0} \|{\bf v}({\bf x}(t),t)\|\ dt \;.
\end{eqnarray}
The forward component, $M^{(f)}$, reveals the stable manifolds, while the backward integration term, $M^{(b)}$, highlights the locations of the unstable manifolds. Our focus in this paper is on the attracting material curves, therefore we will restrict our calculations to $M^{(b)}$, and not on  $M^{(f)}$, which is related to repelling material curves.  
Additionally, \citep{lopesino2017} discusses the conditions under which level curves of this field correspond to invariant sets, in particular tori.  The kinematic model presented in Section \ref{kinematic_model} will serve as a controlled model for illustrating these ideas. Specifically, the level curves of the unperturbed time-independent streamfunction shown in Figure \ref{datakm}e) correspond to closed trajectories of the system described by equation \eqref{sd}. This function acts as the Hamiltonian, as seen in the analogy presented in equation \eqref{velfromtstream}. 
These closed trajectories, associated with invariant tori, act as barriers to transport, preventing mixing on both sides of the jet. Under small time-dependent perturbations, these barriers may be partially broken \cite{kam1, kam2, kam3, Mancho2}. Figure \ref{resultsb}a) displays contour lines of $M^{(b)}$ evaluated for the slightly perturbed kinematic model (i.e., $c_3=0.4$ and $c_4=0.2$) at time $t_0 = 0$ and $\tau = 6600$ s.
 Irregular black contours correspond to singular features in the function, which are, in turn, connected to the presence of invariant unstable manifolds in chaotic regions. Smooth regions indicate the presence of invariant tori. Red and blue circles indicate the initial conditions in both types of domains that do not mix. The results of the ergodic partition theory discussed in \cite{lopesino2017} demonstrate that the converged average of $M^{(b)}/\tau$ as $\tau \to \infty$ has level curves associated with tori.
 These ideas have been applied to identify barriers in the Antarctic polar jet \cite{Curbelo_partI}. The blue line in Figure \ref{resultsb}b) demonstrates good convergence of this average for $\tau>4000$. In contrast, the red line, representing the evolution of the average for an initial condition in a chaotic region as a function of $\tau$, shows that convergence is not achieved within the displayed interval.  A discussion of the lack of convergence in the chaotic setting is provided in \cite{gg2018b}. Larger perturbations may be required in the kinematic model to match the experimental results shown in Figure \ref{kalli2}, which we will now discuss. 

\begin{figure*}[htbp]
\begin{center}
a) \includegraphics[scale=0.2]{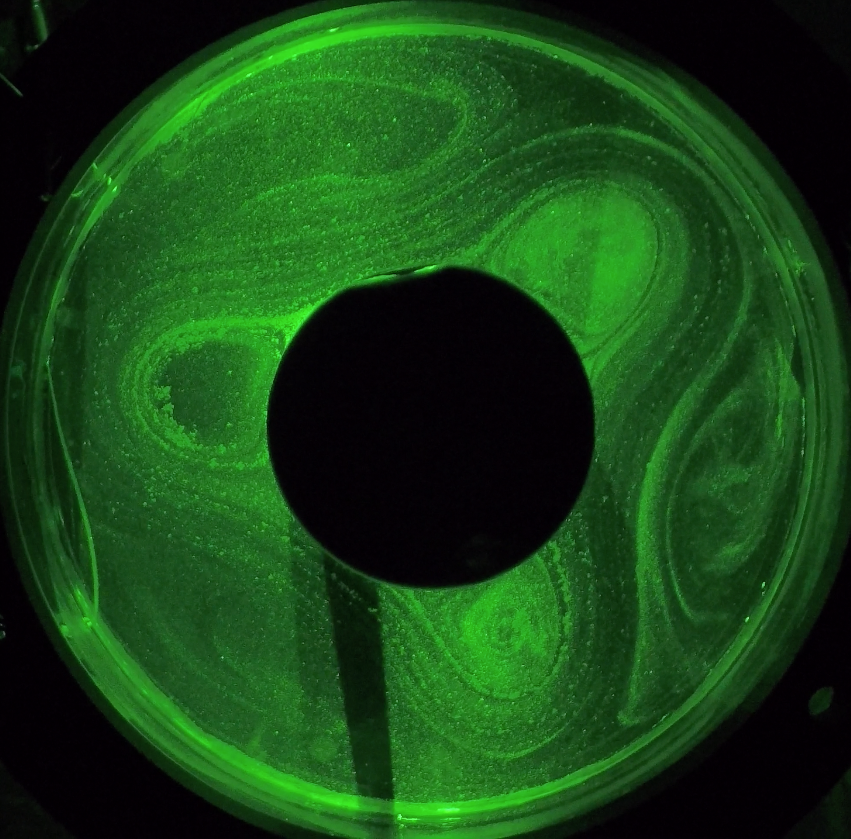}
b) \includegraphics[scale=0.45]{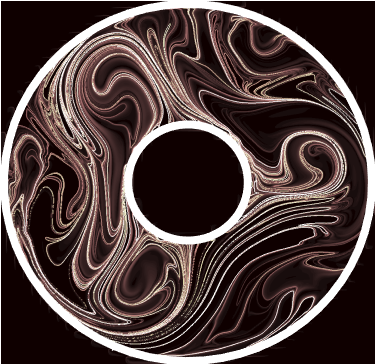}\\
c) \includegraphics[scale=0.45]{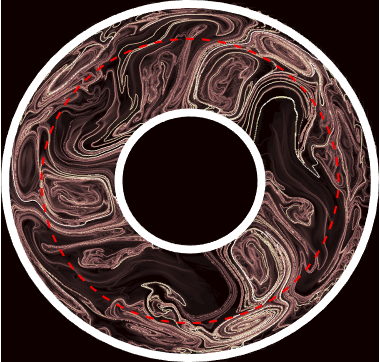}
d) \includegraphics[scale=0.45]{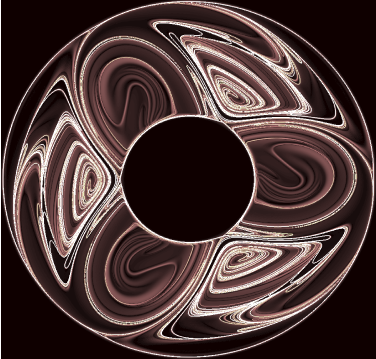}\\
\caption{a) Experimental observation of attracting material curves in the counter-clockwise rotating wave; Panels (b)-(d) show attracting material curves are highlighted using $||\nabla M^{(b)}||$; (b) results using velocities from Subsection III A 1 at the time $t_0$ associated with snapshot 326 and $\tau=250$ s; (c) results  obtained using 
 $\tau = 62.5$ s and velocities obtained from equations \eqref{veltermx}-\eqref{veltermy}. 
 The temperature fields used were obtained through an EOF analysis based on experimental temperatures at the time $t_0$ associated with snapshot 500. The outer red dashed line highlights the limit of boundary effects for the used approximation; d) results computed from the kinematic model at $t_0 = 0$ s and $\tau = 600$s. }
\label{results}
\end{center}
\end{figure*}

Fig. \ref{results} summarizes the comparison between the Lagrangian skeleton obtained  for the three proposed models and the experimental observations. Panel a) displays, for comparison purposes, one of the frames contained in figure \ref{kalli2}. Panels b) to d) highlight  singular features in $M^{(b)}$ computed from the models. In these representations, singular features are accentuated by calculating the modulus of the gradient of $M^{(b)}$, denoted as $|| \nabla M^{(b)} ||$, which exhibits high values along the singular features. While this representation does not allow us to depict tori as contour lines, this is no a critical limitation, as these idealized objects are not present in the experimental setting. Panel b) shows $|| \nabla M^{(b)} ||$ for $M^{(b)}$ computed with $\tau=250$s 
 using the velocities obtained from the model explained in subsection \ref{velocity model}. The time $t_0$ considered is centered around snapshot 326 of the experimental measurements.
 The filamentous structures on the jet, especially in the fourth quadrant, closely resemble those observed in the experiment and within the interior of the clover leaves.
  Panel c) displays $|| \nabla M^{(b)} ||$ for $M^{(b)}$ computed with $\tau = 62.5$ s using the velocities obtained from the model explained in subsection \ref{temperature model}. 
  The temperature data that have been used to estimate the velocity via (\ref{veltermx}) and (\ref{veltermy}) have a higher spatial resolution than the velocity data and hence show finer structures. Moreover, the temperature has been measured at the surface, and the velocity data have been taken from 1cm below the surface. Surface effects related to wind or the ambient temperature and humidity might slightly alter the temperature field and hence the estimated velocity components.
  Note that for panel c), features of anti-cyclonic flow can be observed closer to the inner cylinder than in panel b), corresponding slightly better with the structures visible in Figure \ref{resultsb}a).
  For these data, the time $t_0$ considered for computing $M^{(b)}$ is centered around snapshot 500 of the experimental measurements.
 The outer part of the red circumference delineates a region with spurious structures caused by the outer heating circle. 
 The distortion is a result of the high temperature within the boundary layer along the outer heated boundary, as visible in Figure \ref{veltempeig}d).
    Although this average has been subtracted, an oscillatory pattern in the outer crown of Figure \ref{veltempeig}e) does not appear to correspond to a streamfunction pattern for the velocity. This pattern might be related to a backward bending plume of low temperatures probably related to wind effects in the outer region of the annulus where such effects are most noticeable.
  However, in the inner part of the red circumferences, some similarities with the experiment can be observed in the filamentous structures surrounding the clover leaves. Panel d) displays $|| \nabla M^{(b)} ||$ from an $|| M^{(b)} ||$ computed with $\tau = 600$ s using the kinematic model adjusted for a stronger perturbation  that uses $c_3=2.55$ and $c_4=0.43$. This perturbation allows for mixing across the jet by killing the tori. As a result, the averages of $M^{(b)}/\tau$ do not reach convergence within the tested $\tau$ intervals shown in Figure \ref{resultsb}. 
  In this output, $M^{(b)}$ is computed at time $t_0 = 0$, and both the filamentous structures on the jet and within the interior of the clover leaves closely resemble the experimental observations.  The selected $t_0$ values in the panels have been chosen to better match the experimental observations, but other times may not match as well. We attribute this to potential distortions in the measurements. 
 The fact that different $\tau$ values are required in each model to adjust the observed patterns indicates varying strengths in the attraction of the material curves associated with the vector fields of each model. 

\section{Conclusions}
\label{conclusions}

This work focuses on the study of transport in baroclinic waves from both experimental and modeling perspectives. Lagrangian observations indicated that particles and floats can stay a long time inside of eddies \cite{Provenzale:1999}. It is argued that strong gradients of potential vorticity act as transport barriers \cite{McIntyre:1989}. The baroclinic waves in the experimental setup considered here form a series of baroclinic eddies similar to those found in the mid-latitude atmosphere and it is hence of relevance to understand and simulate the transport of such eddies using the kinematic model approach. In the experimental setting, we have simultaneously measured temperature, velocities, and transport.
We have used the first two types of measurements to construct two models based on EOF and to calibrate a kinematic model. The transport measurements highlight attracting material curves in the underlying system, which is the least noisy observation of all. We have reproduced the attracting material curves for the three models using Lagrangian Descriptors, a technique that allows us to successfully reveal the phase space structures, and found that the models are in good agreement with the transport structures observed in the experiment. In particular, we have shown that the kinematic model and the model obtained from filtered velocities perform well in this regard.

To test the models and the ability to find characteristic structures we applied the Lagrangian descriptor $M$ to experimental data of convective flow in a rather stable  regime. After this first successful application, it is obvious to apply the method to data with more complex spatio-temporal dynamics. Such data correspond to experiments with larger $Ta$ and smaller $Ro$. In that case, the flow does not settle down to a quasi-stationary state but shows vacillations, wave breaking, and vortex shedding. For the future, it will be instructive to apply the trajectory-based diagnostics to such more complex flows to isolate certain mechanisms that lead to the mentioned events as it was already done e.g. in \cite{Madrid}. We finally note that the use of temperature fields to derive velocities is particularly interesting since these simultaneous measurements have allowed for verification and opened up possibilities for applying this methodology to other geostrophic flows, such as in the ocean. In this case, temperature field observations are available, and they could be used to estimate ocean velocities, not only at the surface, but also in the interior \cite{LaCasce:2006}.  

\section{Acknowledgements}
AMM  acknowledges the support of a CSIC PIE project Ref. 202250E001, from grant PID2021-123348OB-I00 funded by 
MCIN/ AEI /10.13039/501100011033/ and by
FEDER A way for making Europe and from ONR Grant N00014-
17-1-3003. AMM is an active member
of the CSIC Interdisciplinary Thematic Platform POLARCSIC and  TELEDETECT.
UH acknowledges financial support from the German Research Foundation DFG
(HA 2932/17-4) and the BTU Graduate Research School. UH further thanks the technicians of the Department of Aerodynamics and Fluid Mechanics at BTU, Robin St\"obel and
Stefan Rohark, for help with the technical equipment, and Franz-Theo Sch{\"o}n for help with the dye visualizations.
MA acknowledges support from the grant CEX2019-000904-S and IJC2019-040168-I funded by: MCIN/AEI/ 10.13039/501100011033  and by “European Union NextGenerationEU/PRTR”.
\bibliography{anystyle}

\begin{thebibliography}{78}%
\makeatletter
\providecommand \@ifxundefined [1]{%
 \@ifx{#1\undefined}
}%
\providecommand \@ifnum [1]{%
 \ifnum #1\expandafter \@firstoftwo
 \else \expandafter \@secondoftwo
 \fi
}%
\providecommand \@ifx [1]{%
 \ifx #1\expandafter \@firstoftwo
 \else \expandafter \@secondoftwo
 \fi
}%
\providecommand \natexlab [1]{#1}%
\providecommand \enquote  [1]{``#1''}%
\providecommand \bibnamefont  [1]{#1}%
\providecommand \bibfnamefont [1]{#1}%
\providecommand \citenamefont [1]{#1}%
\providecommand \href@noop [0]{\@secondoftwo}%
\providecommand \href [0]{\begingroup \@sanitize@url \@href}%
\providecommand \@href[1]{\@@startlink{#1}\@@href}%
\providecommand \@@href[1]{\endgroup#1\@@endlink}%
\providecommand \@sanitize@url [0]{\catcode `\\12\catcode `\$12\catcode
  `\&12\catcode `\#12\catcode `\^12\catcode `\_12\catcode `\%12\relax}%
\providecommand \@@startlink[1]{}%
\providecommand \@@endlink[0]{}%
\providecommand \url  [0]{\begingroup\@sanitize@url \@url }%
\providecommand \@url [1]{\endgroup\@href {#1}{\urlprefix }}%
\providecommand \urlprefix  [0]{URL }%
\providecommand \Eprint [0]{\href }%
\providecommand \doibase [0]{http://dx.doi.org/}%
\providecommand \selectlanguage [0]{\@gobble}%
\providecommand \bibinfo  [0]{\@secondoftwo}%
\providecommand \bibfield  [0]{\@secondoftwo}%
\providecommand \translation [1]{[#1]}%
\providecommand \BibitemOpen [0]{}%
\providecommand \bibitemStop [0]{}%
\providecommand \bibitemNoStop [0]{.\EOS\space}%
\providecommand \EOS [0]{\spacefactor3000\relax}%
\providecommand \BibitemShut  [1]{\csname bibitem#1\endcsname}%
\let\auto@bib@innerbib\@empty
\bibitem [{\citenamefont {Pedlosky}(1989)}]{Pedlosky:89}%
  \BibitemOpen
  \bibfield  {author} {\bibinfo {author} {\bibfnamefont {J.}~\bibnamefont
  {Pedlosky}},\ }\href@noop {} {\emph {\bibinfo {title} {{Geophysical Fluid
  Dynamics}}}}\ (\bibinfo  {publisher} {Springer},\ \bibinfo {year} {1989})\
  p.\ \bibinfo {pages} {710pp.}\BibitemShut {Stop}%
\bibitem [{\citenamefont {Read}\ \emph {et~al.}(2020)\citenamefont {Read},
  \citenamefont {Kennedy}, \citenamefont {Lewis}, \citenamefont {Scolan},
  \citenamefont {Tabataba-Vakili}, \citenamefont {Wang}, \citenamefont
  {Wright},\ and\ \citenamefont {Young}}]{Read_etal:2020}%
  \BibitemOpen
  \bibfield  {author} {\bibinfo {author} {\bibfnamefont {P.}~\bibnamefont
  {Read}}, \bibinfo {author} {\bibfnamefont {D.}~\bibnamefont {Kennedy}},
  \bibinfo {author} {\bibfnamefont {N.}~\bibnamefont {Lewis}}, \bibinfo
  {author} {\bibfnamefont {H.}~\bibnamefont {Scolan}}, \bibinfo {author}
  {\bibfnamefont {F.}~\bibnamefont {Tabataba-Vakili}}, \bibinfo {author}
  {\bibfnamefont {Y.}~\bibnamefont {Wang}}, \bibinfo {author} {\bibfnamefont
  {S.}~\bibnamefont {Wright}}, \ and\ \bibinfo {author} {\bibfnamefont
  {R.}~\bibnamefont {Young}},\ }\bibfield  {title} {\enquote {\bibinfo {title}
  {Baroclinic and barotropic instabilities in planetary atmospheres:
  energetics, equilibration and adjustment},}\ }\href {\doibase
  10.5194/npg-27-147-2020} {\bibfield  {journal} {\bibinfo  {journal}
  {Nonlinear Processes in Geophysics}\ }\textbf {\bibinfo {volume} {27}},\
  \bibinfo {pages} {147--173} (\bibinfo {year} {2020})}\BibitemShut {NoStop}%
\bibitem [{\citenamefont {Fowlis}\ and\ \citenamefont
  {Hide}(1965)}]{Fowlis_and_Hide:65}%
  \BibitemOpen
  \bibfield  {author} {\bibinfo {author} {\bibfnamefont {W.~W.}\ \bibnamefont
  {Fowlis}}\ and\ \bibinfo {author} {\bibfnamefont {R.}~\bibnamefont {Hide}},\
  }\bibfield  {title} {\enquote {\bibinfo {title} {{Thermal convection in a
  rotating annulus of liquid: effect of viscosity on the transition between
  axisymmetric and non-axisymmetric flow regimes}},}\ }\href@noop {} {\bibfield
   {journal} {\bibinfo  {journal} {J. Atmos. Sci.}\ }\textbf {\bibinfo {volume}
  {22}},\ \bibinfo {pages} {541--558} (\bibinfo {year} {1965})}\BibitemShut
  {NoStop}%
\bibitem [{\citenamefont {Read}, \citenamefont {Sommeria},\ and\ \citenamefont
  {Young}(2019)}]{read_sommeria_young_2019}%
  \BibitemOpen
  \bibfield  {author} {\bibinfo {author} {\bibfnamefont {P.~L.}\ \bibnamefont
  {Read}}, \bibinfo {author} {\bibfnamefont {J.}~\bibnamefont {Sommeria}}, \
  and\ \bibinfo {author} {\bibfnamefont {R.~M.}\ \bibnamefont {Young}},\
  }\enquote {\bibinfo {title} {Convectively driven turbulence, rossby waves and
  zonal jets: Experiments on the coriolis platform},}\ in\ \href {\doibase
  10.1017/9781107358225.007} {\emph {\bibinfo {booktitle} {Zonal Jets:
  Phenomenology, Genesis, and Physics}}},\ \bibinfo {editor} {edited by\
  \bibinfo {editor} {\bibfnamefont {B.}~\bibnamefont {Galperin}}\ and\ \bibinfo
  {editor} {\bibfnamefont {P.~L.}\ \bibnamefont {Read}}}\ (\bibinfo
  {publisher} {Cambridge University Press},\ \bibinfo {year} {2019})\ p.\
  \bibinfo {pages} {135–151}\BibitemShut {NoStop}%
\bibitem [{\citenamefont {Morita}(1990)}]{Morita:90}%
  \BibitemOpen
  \bibfield  {author} {\bibinfo {author} {\bibfnamefont {O.}~\bibnamefont
  {Morita}},\ }\bibfield  {title} {\enquote {\bibinfo {title} {{Transition
  between flow regimes of baroclinic flows in a rotating annulus of fluid,
  phase transitions}},}\ }\href@noop {} {\bibfield  {journal} {\bibinfo
  {journal} {J. Atmos. Sci.}\ }\textbf {\bibinfo {volume} {46}},\ \bibinfo
  {pages} {213--244} (\bibinfo {year} {1990})}\BibitemShut {NoStop}%
\bibitem [{\citenamefont {Hart}(1981)}]{Hart:81}%
  \BibitemOpen
  \bibfield  {author} {\bibinfo {author} {\bibfnamefont {J.~E.}\ \bibnamefont
  {Hart}},\ }\bibfield  {title} {\enquote {\bibinfo {title} {Wavenumber
  selection in nonlinear baroclinic instability},}\ }\href {\doibase
  10.1175/1520-0469(1981)038<0400:WSINBI>2.0.CO;2} {\bibfield  {journal}
  {\bibinfo  {journal} {J. Atmos. Sci.}\ }\textbf {\bibinfo {volume} {38}},\
  \bibinfo {pages} {400 -- 408} (\bibinfo {year} {1981})}\BibitemShut {NoStop}%
\bibitem [{\citenamefont {Fr{\"u}h}\ and\ \citenamefont
  {Read}(1997)}]{Frueh_and_Read:97}%
  \BibitemOpen
  \bibfield  {author} {\bibinfo {author} {\bibfnamefont {W.~G.}\ \bibnamefont
  {Fr{\"u}h}}\ and\ \bibinfo {author} {\bibfnamefont {P.~L.}\ \bibnamefont
  {Read}},\ }\bibfield  {title} {\enquote {\bibinfo {title} {{Wave interactions
  and the transition to chaos of baroclinic waves in a thermally driven
  rotating annulus}},}\ }\href@noop {} {\bibfield  {journal} {\bibinfo
  {journal} {Phil. Trans. Roy. Soc. London}\ }\textbf {\bibinfo {volume}
  {A355}},\ \bibinfo {pages} {101--153} (\bibinfo {year} {1997})}\BibitemShut
  {NoStop}%
\bibitem [{\citenamefont {Lindzen}, \citenamefont {Farrell},\ and\
  \citenamefont {Jacqmin}(1981)}]{Lindzen_etal:81}%
  \BibitemOpen
  \bibfield  {author} {\bibinfo {author} {\bibfnamefont {R.~S.}\ \bibnamefont
  {Lindzen}}, \bibinfo {author} {\bibfnamefont {B.}~\bibnamefont {Farrell}}, \
  and\ \bibinfo {author} {\bibfnamefont {D.}~\bibnamefont {Jacqmin}},\
  }\bibfield  {title} {\enquote {\bibinfo {title} {{Vascillations due to wave
  interference: application to the atmosphere and to annulus experiments}},}\
  }\href@noop {} {\bibfield  {journal} {\bibinfo  {journal} {J. Atmos. Sci.}\
  }\textbf {\bibinfo {volume} {39}},\ \bibinfo {pages} {14--23} (\bibinfo
  {year} {1981})}\BibitemShut {NoStop}%
\bibitem [{\citenamefont {Rodda}\ and\ \citenamefont
  {Harlander}(2020)}]{Rodda_and_Harlander:20}%
  \BibitemOpen
  \bibfield  {author} {\bibinfo {author} {\bibfnamefont {C.}~\bibnamefont
  {Rodda}}\ and\ \bibinfo {author} {\bibfnamefont {U.}~\bibnamefont
  {Harlander}},\ }\bibfield  {title} {\enquote {\bibinfo {title} {{Transition
  from geostrophic flows to inertia-gravity waves in the spectrum of a
  differentially heated rotating annulus experiment}},}\ }\href@noop {}
  {\bibfield  {journal} {\bibinfo  {journal} {J. Atmos. Sci.}\ }\textbf
  {\bibinfo {volume} {77(8)}},\ \bibinfo {pages} {2793--2806} (\bibinfo {year}
  {2020})}\BibitemShut {NoStop}%
\bibitem [{\citenamefont {Vincze}\ \emph {et~al.}(2014)\citenamefont {Vincze},
  \citenamefont {Harlander}, \citenamefont {von Larcher},\ and\ \citenamefont
  {Egbers}}]{Vincze_etal:2014}%
  \BibitemOpen
  \bibfield  {author} {\bibinfo {author} {\bibfnamefont {M.}~\bibnamefont
  {Vincze}}, \bibinfo {author} {\bibfnamefont {U.}~\bibnamefont {Harlander}},
  \bibinfo {author} {\bibfnamefont {T.}~\bibnamefont {von Larcher}}, \ and\
  \bibinfo {author} {\bibfnamefont {C.}~\bibnamefont {Egbers}},\ }\bibfield
  {title} {\enquote {\bibinfo {title} {An experimental study of regime
  transitions in a differentially heated baroclinic annulus with flat and
  sloping bottom topographies},}\ }\href {\doibase 10.5194/npg-21-237-2014}
  {\bibfield  {journal} {\bibinfo  {journal} {Nonlinear Processes in
  Geophysics}\ }\textbf {\bibinfo {volume} {21}},\ \bibinfo {pages} {237--250}
  (\bibinfo {year} {2014})}\BibitemShut {NoStop}%
\bibitem [{\citenamefont {Read}\ \emph {et~al.}(2014)\citenamefont {Read},
  \citenamefont {P{\'e}rez}, \citenamefont {Moroz},\ and\ \citenamefont
  {Young}}]{Read_etal:2014}%
  \BibitemOpen
  \bibfield  {author} {\bibinfo {author} {\bibfnamefont {P.~L.}\ \bibnamefont
  {Read}}, \bibinfo {author} {\bibfnamefont {E.~P.}\ \bibnamefont {P{\'e}rez}},
  \bibinfo {author} {\bibfnamefont {I.~M.}\ \bibnamefont {Moroz}}, \ and\
  \bibinfo {author} {\bibfnamefont {R.~M.~B.}\ \bibnamefont {Young}},\
  }\enquote {\bibinfo {title} {General circulation of planetary atmospheres:
  Insights from rotating annulus and related experiments.}}\ in\ \href@noop {}
  {\emph {\bibinfo {booktitle} {Modelling Atmospheric and Oceanic Flows:
  Insights from Laboratory Experiments and Numerical Simulations}}},\ \bibinfo
  {editor} {edited by\ \bibinfo {editor} {\bibfnamefont {T.}~\bibnamefont {von
  Larcher}}\ and\ \bibinfo {editor} {\bibfnamefont {P.~D.}\ \bibnamefont
  {Williams}}}\ (\bibinfo  {publisher} {American Geophysical Union},\ \bibinfo
  {address} {Boston, USA},\ \bibinfo {year} {2014})\ pp.\ \bibinfo {pages}
  {9--44}\BibitemShut {NoStop}%
\bibitem [{\citenamefont {Sommeria}, \citenamefont {Meyers},\ and\
  \citenamefont {Swinney}(1989)}]{Sommeria1}%
  \BibitemOpen
  \bibfield  {author} {\bibinfo {author} {\bibfnamefont {J.}~\bibnamefont
  {Sommeria}}, \bibinfo {author} {\bibfnamefont {S.~D.}\ \bibnamefont
  {Meyers}}, \ and\ \bibinfo {author} {\bibfnamefont {H.~L.}\ \bibnamefont
  {Swinney}},\ }\bibfield  {title} {\enquote {\bibinfo {title} {Laboratory
  model of a planetary eastward jet},}\ }\href {\doibase 10.1038/337058a0}
  {\bibfield  {journal} {\bibinfo  {journal} {Nature}\ }\textbf {\bibinfo
  {volume} {337}},\ \bibinfo {pages} {58--61} (\bibinfo {year}
  {1989})}\BibitemShut {NoStop}%
\bibitem [{\citenamefont {Keane}, \citenamefont {Read},\ and\ \citenamefont
  {King}(2014)}]{Keane_etal14}%
  \BibitemOpen
  \bibfield  {author} {\bibinfo {author} {\bibfnamefont {R.~J.}\ \bibnamefont
  {Keane}}, \bibinfo {author} {\bibfnamefont {P.~L.}\ \bibnamefont {Read}}, \
  and\ \bibinfo {author} {\bibfnamefont {G.~P.}\ \bibnamefont {King}},\
  }\bibfield  {title} {\enquote {\bibinfo {title} {{On the stirring properties
  of the thermally-driven rotating annulus}},}\ }\href@noop {} {\bibfield
  {journal} {\bibinfo  {journal} {Physica D}\ }\textbf {\bibinfo {volume}
  {268}},\ \bibinfo {pages} {50--58} (\bibinfo {year} {2014})}\BibitemShut
  {NoStop}%
\bibitem [{\citenamefont {Joseph}\ and\ \citenamefont
  {Legras}(2002)}]{bernard}%
  \BibitemOpen
  \bibfield  {author} {\bibinfo {author} {\bibfnamefont {B.}~\bibnamefont
  {Joseph}}\ and\ \bibinfo {author} {\bibfnamefont {B.}~\bibnamefont
  {Legras}},\ }\bibfield  {title} {\enquote {\bibinfo {title} {Relation between
  kinematic boundaries, stirring, and barriers for the {A}ntarctic polar
  vortex},}\ }\href@noop {} {\bibfield  {journal} {\bibinfo  {journal} {J.
  Atmos. Sci.}\ }\textbf {\bibinfo {volume} {59}},\ \bibinfo {pages}
  {1198--1212} (\bibinfo {year} {2002})}\BibitemShut {NoStop}%
\bibitem [{\citenamefont {de~la C\'amara}\ \emph {et~al.}(2010)\citenamefont
  {de~la C\'amara}, \citenamefont {Mechoso}, \citenamefont {Ide}, \citenamefont
  {Walterscheid},\ and\ \citenamefont {Schubert}}]{alvaro10}%
  \BibitemOpen
  \bibfield  {author} {\bibinfo {author} {\bibfnamefont {A.}~\bibnamefont
  {de~la C\'amara}}, \bibinfo {author} {\bibfnamefont {C.~R.}\ \bibnamefont
  {Mechoso}}, \bibinfo {author} {\bibfnamefont {K.}~\bibnamefont {Ide}},
  \bibinfo {author} {\bibfnamefont {R.}~\bibnamefont {Walterscheid}}, \ and\
  \bibinfo {author} {\bibfnamefont {G.}~\bibnamefont {Schubert}},\ }\bibfield
  {title} {\enquote {\bibinfo {title} {Polar night vortex breakdown and
  large-scale stirring in the southern stratosphere},}\ }\href@noop {}
  {\bibfield  {journal} {\bibinfo  {journal} {Climate Dynamics}\ }\textbf
  {\bibinfo {volume} {35}},\ \bibinfo {pages} {965--975} (\bibinfo {year}
  {2010})}\BibitemShut {NoStop}%
\bibitem [{\citenamefont {de~la C{\'a}mara}\ \emph
  {et~al.}(2012{\natexlab{a}})\citenamefont {de~la C{\'a}mara}, \citenamefont
  {Mancho}, \citenamefont {Ide}, \citenamefont {Serrano},\ and\ \citenamefont
  {Mechoso}}]{alvaro1}%
  \BibitemOpen
  \bibfield  {author} {\bibinfo {author} {\bibfnamefont {A.}~\bibnamefont
  {de~la C{\'a}mara}}, \bibinfo {author} {\bibfnamefont {A.~M.}\ \bibnamefont
  {Mancho}}, \bibinfo {author} {\bibfnamefont {K.}~\bibnamefont {Ide}},
  \bibinfo {author} {\bibfnamefont {E.}~\bibnamefont {Serrano}}, \ and\
  \bibinfo {author} {\bibfnamefont {C.}~\bibnamefont {Mechoso}},\ }\bibfield
  {title} {\enquote {\bibinfo {title} {{Routes of transport across the
  Antarctic polar vortex in the southern spring.}}}\ }\href@noop {} {\bibfield
  {journal} {\bibinfo  {journal} {J. Atmos. Sci.}\ }\textbf {\bibinfo {volume}
  {69}},\ \bibinfo {pages} {753--767} (\bibinfo {year}
  {2012}{\natexlab{a}})}\BibitemShut {NoStop}%
\bibitem [{\citenamefont {de~la C{\'a}mara}\ \emph {et~al.}(2013)\citenamefont
  {de~la C{\'a}mara}, \citenamefont {Mechoso}, \citenamefont {Serrano},\ and\
  \citenamefont {Ide}}]{alvaro2}%
  \BibitemOpen
  \bibfield  {author} {\bibinfo {author} {\bibfnamefont {A.}~\bibnamefont
  {de~la C{\'a}mara}}, \bibinfo {author} {\bibfnamefont {C.~R.}\ \bibnamefont
  {Mechoso}}, \bibinfo {author} {\bibfnamefont {E.}~\bibnamefont {Serrano}}, \
  and\ \bibinfo {author} {\bibfnamefont {K.}~\bibnamefont {Ide}},\ }\bibfield
  {title} {\enquote {\bibinfo {title} {{Quasi-horizontal transport within the
  Antarctic polar night vortex: Rossby wave breaking evidence and Lagrangian
  structures}},}\ }\href@noop {} {\bibfield  {journal} {\bibinfo  {journal} {J.
  Atmos. Sci.}\ }\textbf {\bibinfo {volume} {70}},\ \bibinfo {pages}
  {2982--3001} (\bibinfo {year} {2013})}\BibitemShut {NoStop}%
\bibitem [{\citenamefont {Garc\'{\i}a-Garrido}\ \emph
  {et~al.}(2017)\citenamefont {Garc\'{\i}a-Garrido}, \citenamefont {Curbelo},
  \citenamefont {Mechoso}, \citenamefont {Mancho},\ and\ \citenamefont
  {Wiggins}}]{Garcia2017}%
  \BibitemOpen
  \bibfield  {author} {\bibinfo {author} {\bibfnamefont {V.~J.}\ \bibnamefont
  {Garc\'{\i}a-Garrido}}, \bibinfo {author} {\bibfnamefont {J.}~\bibnamefont
  {Curbelo}}, \bibinfo {author} {\bibfnamefont {C.~R.}\ \bibnamefont
  {Mechoso}}, \bibinfo {author} {\bibfnamefont {A.~M.}\ \bibnamefont {Mancho}},
  \ and\ \bibinfo {author} {\bibfnamefont {S.}~\bibnamefont {Wiggins}},\
  }\bibfield  {title} {\enquote {\bibinfo {title} {A simple kinematic model for
  the lagrangian description of relevant nonlinear processes in the
  stratospheric polar vortex},}\ }\href {\doibase 10.5194/npg-24-265-2017}
  {\bibfield  {journal} {\bibinfo  {journal} {Nonlinear Processes in
  Geophysics}\ }\textbf {\bibinfo {volume} {24}},\ \bibinfo {pages} {265--278}
  (\bibinfo {year} {2017})}\BibitemShut {NoStop}%
\bibitem [{\citenamefont {Curbelo}\ \emph {et~al.}(2017)\citenamefont
  {Curbelo}, \citenamefont {Garc\'{i}-Garrido}, \citenamefont {Mechoso},
  \citenamefont {Mancho}, \citenamefont {Wiggins},\ and\ \citenamefont
  {Niang}}]{Curbelo}%
  \BibitemOpen
  \bibfield  {author} {\bibinfo {author} {\bibfnamefont {J.}~\bibnamefont
  {Curbelo}}, \bibinfo {author} {\bibfnamefont {V.~J.}\ \bibnamefont
  {Garc\'{i}-Garrido}}, \bibinfo {author} {\bibfnamefont {C.~R.}\ \bibnamefont
  {Mechoso}}, \bibinfo {author} {\bibfnamefont {A.~M.}\ \bibnamefont {Mancho}},
  \bibinfo {author} {\bibfnamefont {S.}~\bibnamefont {Wiggins}}, \ and\
  \bibinfo {author} {\bibfnamefont {C.}~\bibnamefont {Niang}},\ }\bibfield
  {title} {\enquote {\bibinfo {title} {Insights into the three-dimensional
  lagrangian geometry of the antarctic polar vortex},}\ }\href {\doibase
  10.5194/npg-24-379-2017} {\bibfield  {journal} {\bibinfo  {journal}
  {Nonlinear Processes in Geophysics}\ }\textbf {\bibinfo {volume} {24}},\
  \bibinfo {pages} {379--392} (\bibinfo {year} {2017})}\BibitemShut {NoStop}%
\bibitem [{\citenamefont {Curbelo}\ \emph
  {et~al.}(2019{\natexlab{a}})\citenamefont {Curbelo}, \citenamefont {Mechoso},
  \citenamefont {Mancho},\ and\ \citenamefont {et~al.}}]{Curbelo_partI}%
  \BibitemOpen
  \bibfield  {author} {\bibinfo {author} {\bibfnamefont {J.}~\bibnamefont
  {Curbelo}}, \bibinfo {author} {\bibfnamefont {C.}~\bibnamefont {Mechoso}},
  \bibinfo {author} {\bibfnamefont {A.}~\bibnamefont {Mancho}}, \ and\ \bibinfo
  {author} {\bibnamefont {et~al.}},\ }\bibfield  {title} {\enquote {\bibinfo
  {title} {Lagrangian study of the final warming in the southern stratosphere
  during 2002: Part i. the vortex splitting at upper levels},}\ }\href
  {\doibase https://doi.org/10.1007/s00382-019-04832-y} {\bibfield  {journal}
  {\bibinfo  {journal} {Clim. Dyn.}\ }\textbf {\bibinfo {volume} {53}},\
  \bibinfo {pages} {2779--2792} (\bibinfo {year}
  {2019}{\natexlab{a}})}\BibitemShut {NoStop}%
\bibitem [{\citenamefont {Curbelo}\ \emph
  {et~al.}(2019{\natexlab{b}})\citenamefont {Curbelo}, \citenamefont {Mechoso},
  \citenamefont {Mancho},\ and\ \citenamefont {et~al.}}]{Curbelo_partII}%
  \BibitemOpen
  \bibfield  {author} {\bibinfo {author} {\bibfnamefont {J.}~\bibnamefont
  {Curbelo}}, \bibinfo {author} {\bibfnamefont {C.}~\bibnamefont {Mechoso}},
  \bibinfo {author} {\bibfnamefont {A.}~\bibnamefont {Mancho}}, \ and\ \bibinfo
  {author} {\bibnamefont {et~al.}},\ }\bibfield  {title} {\enquote {\bibinfo
  {title} {Lagrangian study of the final warming in the southern stratosphere
  during 2002: Part ii. 3d structure},}\ }\href {\doibase
  https://doi.org/10.1007/s00382-019-04833-x} {\bibfield  {journal} {\bibinfo
  {journal} {Clim. Dyn.}\ }\textbf {\bibinfo {volume} {53}},\ \bibinfo {pages}
  {1277--1286} (\bibinfo {year} {2019}{\natexlab{b}})}\BibitemShut {NoStop}%
\bibitem [{\citenamefont {Behringer}, \citenamefont {Meyers},\ and\
  \citenamefont {Swinney}(1991)}]{behringer1991}%
  \BibitemOpen
  \bibfield  {author} {\bibinfo {author} {\bibfnamefont {R.~P.}\ \bibnamefont
  {Behringer}}, \bibinfo {author} {\bibfnamefont {S.~D.}\ \bibnamefont
  {Meyers}}, \ and\ \bibinfo {author} {\bibfnamefont {H.~L.}\ \bibnamefont
  {Swinney}},\ }\bibfield  {title} {\enquote {\bibinfo {title} {Chaos and
  mixing in a geostrophic flow},}\ }\href {\doibase 10.1063/1.858052}
  {\bibfield  {journal} {\bibinfo  {journal} {Physics of Fluids A: Fluid
  Dynamics}\ }\textbf {\bibinfo {volume} {3}},\ \bibinfo {pages} {1243--1249}
  (\bibinfo {year} {1991})}\BibitemShut {NoStop}%
\bibitem [{\citenamefont {Scolan}\ and\ \citenamefont
  {Read}(2017)}]{Scolan_etal:2017}%
  \BibitemOpen
  \bibfield  {author} {\bibinfo {author} {\bibfnamefont {H.}~\bibnamefont
  {Scolan}}\ and\ \bibinfo {author} {\bibfnamefont {P.}~\bibnamefont {Read}},\
  }\bibfield  {title} {\enquote {\bibinfo {title} {A rotating annulus driven by
  localized convective forcing: a new atmosphere-like experiment},}\ }\href
  {\doibase 10.1007/s00348-017-2347-5} {\bibfield  {journal} {\bibinfo
  {journal} {Exp. Fluids}\ }\textbf {\bibinfo {volume} {58}} (\bibinfo {year}
  {2017}),\ 10.1007/s00348-017-2347-5}\BibitemShut {NoStop}%
\bibitem [{\citenamefont {Swarnakar}, \citenamefont {Bhattacharya},\ and\
  \citenamefont {Balasubramanian}(2023)}]{Svarnakar:2023}%
  \BibitemOpen
  \bibfield  {author} {\bibinfo {author} {\bibfnamefont {S.}~\bibnamefont
  {Swarnakar}}, \bibinfo {author} {\bibfnamefont {A.}~\bibnamefont
  {Bhattacharya}}, \ and\ \bibinfo {author} {\bibfnamefont {S.}~\bibnamefont
  {Balasubramanian}},\ }\bibfield  {title} {\enquote {\bibinfo {title}
  {{Numerical study of rotating convection with bi-directional temperature
  gradients}},}\ }\href {\doibase 10.1063/5.0145965} {\bibfield  {journal}
  {\bibinfo  {journal} {Physics of Fluids}\ }\textbf {\bibinfo {volume} {35}},\
  \bibinfo {pages} {056601} (\bibinfo {year} {2023})},\ \Eprint
  {http://arxiv.org/abs/https://pubs.aip.org/aip/pof/article-pdf/doi/10.1063/5.0145965/17275489/056601\_1\_5.0145965.pdf}
  {https://pubs.aip.org/aip/pof/article-pdf/doi/10.1063/5.0145965/17275489/056601\_1\_5.0145965.pdf}
  \BibitemShut {NoStop}%
\bibitem [{\citenamefont {Garc\'ia-Garrido}\ \emph {et~al.}(2017)\citenamefont
  {Garc\'ia-Garrido}, \citenamefont {Curbelo}, \citenamefont {Mechoso},
  \citenamefont {Mancho},\ and\ \citenamefont {Wiggins}}]{AGP}%
  \BibitemOpen
  \bibfield  {author} {\bibinfo {author} {\bibfnamefont {V.~J.}\ \bibnamefont
  {Garc\'ia-Garrido}}, \bibinfo {author} {\bibfnamefont {J.}~\bibnamefont
  {Curbelo}}, \bibinfo {author} {\bibfnamefont {C.~R.}\ \bibnamefont
  {Mechoso}}, \bibinfo {author} {\bibfnamefont {A.~M.}\ \bibnamefont {Mancho}},
  \ and\ \bibinfo {author} {\bibfnamefont {S.}~\bibnamefont {Wiggins}},\
  }\bibfield  {title} {\enquote {\bibinfo {title} {A simple kinematic model for
  the lagrangian description of relevant nonlinear processes in the
  stratospheric polar vortex},}\ }\href {\doibase 10.5194/npg-24-265-2017}
  {\bibfield  {journal} {\bibinfo  {journal} {Nonlinear Processes Geophysics}\
  }\textbf {\bibinfo {volume} {24}} (\bibinfo {year} {2017}),\
  10.5194/npg-24-265-2017}\BibitemShut {NoStop}%
\bibitem [{\citenamefont {del Castillo-Negrete}\ and\ \citenamefont
  {Morrison}(1993)}]{Castillo}%
  \BibitemOpen
  \bibfield  {author} {\bibinfo {author} {\bibfnamefont {D.}~\bibnamefont {del
  Castillo-Negrete}}\ and\ \bibinfo {author} {\bibfnamefont {P.~J.}\
  \bibnamefont {Morrison}},\ }\bibfield  {title} {\enquote {\bibinfo {title}
  {Chaotic transport by rossby waves in shear flow},}\ }\href {\doibase
  10.1063/1.858639} {\bibfield  {journal} {\bibinfo  {journal} {Physics of
  Fluids A: Fluid Dynamics}\ }\textbf {\bibinfo {volume} {5}} (\bibinfo {year}
  {1993}),\ 10.1063/1.858639}\BibitemShut {NoStop}%
\bibitem [{\citenamefont {Samelson}(1992)}]{Samelson}%
  \BibitemOpen
  \bibfield  {author} {\bibinfo {author} {\bibfnamefont {R.~M.}\ \bibnamefont
  {Samelson}},\ }\bibfield  {title} {\enquote {\bibinfo {title} {Fluid exchange
  across a meandering jet},}\ }\href {\doibase
  10.1175/1520-0485(1992)022<0431:FEAAMJ>2.0.CO;2} {\bibfield  {journal}
  {\bibinfo  {journal} {Journal of Physical Oceanography}\ }\textbf {\bibinfo
  {volume} {22}} (\bibinfo {year} {1992}),\
  10.1175/1520-0485(1992)022<0431:FEAAMJ>2.0.CO;2}\BibitemShut {NoStop}%
\bibitem [{\citenamefont {Samelson}\ and\ \citenamefont
  {Wiggins}(2006)}]{Wiggins}%
  \BibitemOpen
  \bibfield  {author} {\bibinfo {author} {\bibfnamefont {R.~M.}\ \bibnamefont
  {Samelson}}\ and\ \bibinfo {author} {\bibfnamefont {S.}~\bibnamefont
  {Wiggins}},\ }\href@noop {} {\emph {\bibinfo {title} {Lagrangian Transport in
  Geophysical Jets and Waves: The Dynamical Systems Approach}}}\ (\bibinfo
  {publisher} {Springer},\ \bibinfo {year} {2006})\BibitemShut {NoStop}%
\bibitem [{\citenamefont {Wiggins}\ and\ \citenamefont
  {Mancho}(2014)}]{Mancho2}%
  \BibitemOpen
  \bibfield  {author} {\bibinfo {author} {\bibfnamefont {S.}~\bibnamefont
  {Wiggins}}\ and\ \bibinfo {author} {\bibfnamefont {A.~M.}\ \bibnamefont
  {Mancho}},\ }\bibfield  {title} {\enquote {\bibinfo {title} {{Barriers to
  transport in aperiodically time-dependent two dimensional velocity fields:
  Nekhoroshev's Theorem and 'Nearly Invariant' Tori}},}\ }\href {\doibase
  10.5194/npg-21-165-2014} {\bibfield  {journal} {\bibinfo  {journal}
  {Nonlinear Processes in Geophysics}\ }\textbf {\bibinfo {volume} {21}}
  (\bibinfo {year} {2014}),\ 10.5194/npg-21-165-2014}\BibitemShut {NoStop}%
\bibitem [{\citenamefont {Sommeria}, \citenamefont {Meyers},\ and\
  \citenamefont {Swinney}(1991)}]{Sommeria2}%
  \BibitemOpen
  \bibfield  {author} {\bibinfo {author} {\bibfnamefont {J.}~\bibnamefont
  {Sommeria}}, \bibinfo {author} {\bibfnamefont {S.~D.}\ \bibnamefont
  {Meyers}}, \ and\ \bibinfo {author} {\bibfnamefont {H.~L.}\ \bibnamefont
  {Swinney}},\ }\href@noop {} {\enquote {\bibinfo {title} {Experiments on
  vortices and rossby waves in eastward and westward jets},}\ } (\bibinfo
  {year} {1991})\BibitemShut {NoStop}%
\bibitem [{\citenamefont {Uhl}, \citenamefont {Friedrich},\ and\ \citenamefont
  {Haken}(1993)}]{uhl}%
  \BibitemOpen
  \bibfield  {author} {\bibinfo {author} {\bibfnamefont {C.}~\bibnamefont
  {Uhl}}, \bibinfo {author} {\bibfnamefont {R.}~\bibnamefont {Friedrich}}, \
  and\ \bibinfo {author} {\bibfnamefont {H.}~\bibnamefont {Haken}},\ }\bibfield
   {title} {\enquote {\bibinfo {title} {Reconstruction of spatio-temporal
  signals of complex systems},}\ }\href {\doibase 10.1007/BF01312180}
  {\bibfield  {journal} {\bibinfo  {journal} {Zeitschrift f{\"u}r Physik B:
  Condensed Matter}\ }\textbf {\bibinfo {volume} {92}} (\bibinfo {year}
  {1993}),\ 10.1007/BF01312180}\BibitemShut {NoStop}%
\bibitem [{\citenamefont {Kutz}\ \emph {et~al.}(2016)\citenamefont {Kutz},
  \citenamefont {Brunton}, \citenamefont {Brunton},\ and\ \citenamefont
  {Proctor}}]{kut}%
  \BibitemOpen
  \bibfield  {author} {\bibinfo {author} {\bibfnamefont {J.~N.}\ \bibnamefont
  {Kutz}}, \bibinfo {author} {\bibfnamefont {S.~L.}\ \bibnamefont {Brunton}},
  \bibinfo {author} {\bibfnamefont {B.~W.}\ \bibnamefont {Brunton}}, \ and\
  \bibinfo {author} {\bibfnamefont {J.~L.}\ \bibnamefont {Proctor}},\
  }\href@noop {} {\emph {\bibinfo {title} {{Dynamic Mode Decomposition.
  Data-Driven Modeling of Complex Systems}}}}\ (\bibinfo  {publisher} {SIAM},\
  \bibinfo {year} {2016})\BibitemShut {NoStop}%
\bibitem [{\citenamefont {Mezic}(2005)}]{mezic1}%
  \BibitemOpen
  \bibfield  {author} {\bibinfo {author} {\bibfnamefont {I.}~\bibnamefont
  {Mezic}},\ }\bibfield  {title} {\enquote {\bibinfo {title} {Spectral
  properties of dynamical systems, model reduction and decompositions.}}\
  }\href@noop {} {\bibfield  {journal} {\bibinfo  {journal} {Nonlinear
  Dynamics}\ }\textbf {\bibinfo {volume} {41}},\ \bibinfo {pages} {309–325}
  (\bibinfo {year} {2005})}\BibitemShut {NoStop}%
\bibitem [{\citenamefont {Mezic}(2013)}]{mezic2}%
  \BibitemOpen
  \bibfield  {author} {\bibinfo {author} {\bibfnamefont {I.}~\bibnamefont
  {Mezic}},\ }\bibfield  {title} {\enquote {\bibinfo {title} {Analysis of fluid
  flows via spectral properties of the koopman operator.}}\ }\href@noop {}
  {\bibfield  {journal} {\bibinfo  {journal} {Annual Review of Fluid
  Mechanics}\ }\textbf {\bibinfo {volume} {45}},\ \bibinfo {pages} {57–378}
  (\bibinfo {year} {2013})}\BibitemShut {NoStop}%
\bibitem [{\citenamefont {Fein}(1973)}]{Fein:1973}%
  \BibitemOpen
  \bibfield  {author} {\bibinfo {author} {\bibfnamefont {J.~S.}\ \bibnamefont
  {Fein}},\ }\bibfield  {title} {\enquote {\bibinfo {title} {An experimental
  study of the effects of the upper boundary condition on the thermal
  convection in a rotating, differentially heated cylindrical annulus of
  water},}\ }\href@noop {} {\bibfield  {journal} {\bibinfo  {journal}
  {Geophysical Fluid Dynamics}\ }\textbf {\bibinfo {volume} {5}},\ \bibinfo
  {pages} {213--248} (\bibinfo {year} {1973})}\BibitemShut {NoStop}%
\bibitem [{\citenamefont {Andrei~Sukhanovskii}\ and\ \citenamefont
  {Vasiliev}(2023)}]{Sukha:2023}%
  \BibitemOpen
  \bibfield  {author} {\bibinfo {author} {\bibfnamefont {E.~P.}\ \bibnamefont
  {Andrei~Sukhanovskii}}\ and\ \bibinfo {author} {\bibfnamefont
  {A.}~\bibnamefont {Vasiliev}},\ }\bibfield  {title} {\enquote {\bibinfo
  {title} {A shallow layer laboratory model of large-scale atmospheric
  circulation},}\ }\href {\doibase 10.1080/03091929.2023.2220877} {\bibfield
  {journal} {\bibinfo  {journal} {Geophysical \& Astrophysical Fluid Dynamics}\
  }\textbf {\bibinfo {volume} {117}},\ \bibinfo {pages} {155--176} (\bibinfo
  {year} {2023})},\ \Eprint
  {http://arxiv.org/abs/https://doi.org/10.1080/03091929.2023.2220877}
  {https://doi.org/10.1080/03091929.2023.2220877} \BibitemShut {NoStop}%
\bibitem [{\citenamefont {Rodda}\ \emph {et~al.}(2020)\citenamefont {Rodda},
  \citenamefont {Hien}, \citenamefont {Achatz},\ and\ \citenamefont
  {Harlander}}]{Rodda_etal:2020}%
  \BibitemOpen
  \bibfield  {author} {\bibinfo {author} {\bibfnamefont {C.}~\bibnamefont
  {Rodda}}, \bibinfo {author} {\bibfnamefont {S.}~\bibnamefont {Hien}},
  \bibinfo {author} {\bibfnamefont {U.}~\bibnamefont {Achatz}}, \ and\ \bibinfo
  {author} {\bibfnamefont {U.}~\bibnamefont {Harlander}},\ }\bibfield  {title}
  {\enquote {\bibinfo {title} {A new atmospheric-like differentially heated
  rotating annulus configuration to study gravity wave emission from jets and
  fronts},}\ }\href {\doibase 10.1007/s00348-019-2825-z} {\bibfield  {journal}
  {\bibinfo  {journal} {Exp. Fluids}\ }\textbf {\bibinfo {volume} {61}}
  (\bibinfo {year} {2020}),\ 10.1007/s00348-019-2825-z}\BibitemShut {NoStop}%
\bibitem [{\citenamefont {Harlander}\ \emph {et~al.}(2023)\citenamefont
  {Harlander}, \citenamefont {Sukhanovskii}, \citenamefont {Abide},
  \citenamefont {Borcia}, \citenamefont {Popova}, \citenamefont {Rodda},
  \citenamefont {Vasiliev},\ and\ \citenamefont
  {Vincze}}]{Harlander_etal_atmos23}%
  \BibitemOpen
  \bibfield  {author} {\bibinfo {author} {\bibfnamefont {U.}~\bibnamefont
  {Harlander}}, \bibinfo {author} {\bibfnamefont {A.}~\bibnamefont
  {Sukhanovskii}}, \bibinfo {author} {\bibfnamefont {S.}~\bibnamefont {Abide}},
  \bibinfo {author} {\bibfnamefont {I.~D.}\ \bibnamefont {Borcia}}, \bibinfo
  {author} {\bibfnamefont {E.}~\bibnamefont {Popova}}, \bibinfo {author}
  {\bibfnamefont {C.}~\bibnamefont {Rodda}}, \bibinfo {author} {\bibfnamefont
  {A.}~\bibnamefont {Vasiliev}}, \ and\ \bibinfo {author} {\bibfnamefont
  {M.}~\bibnamefont {Vincze}},\ }\bibfield  {title} {\enquote {\bibinfo {title}
  {New laboratory experiments to study the large-scale circulation and climate
  dynamics},}\ }\href {\doibase 10.3390/atmos14050836} {\bibfield  {journal}
  {\bibinfo  {journal} {Atmosphere}\ }\textbf {\bibinfo {volume} {14}}
  (\bibinfo {year} {2023}),\ 10.3390/atmos14050836}\BibitemShut {NoStop}%
\bibitem [{\citenamefont {Haken}(1983)}]{haken}%
  \BibitemOpen
  \bibfield  {author} {\bibinfo {author} {\bibfnamefont {H.}~\bibnamefont
  {Haken}},\ }\href@noop {} {\emph {\bibinfo {title} {{Synergetics, an
  Introduction: Nonequilibrium Phase Transitions and Self-Organization in
  Physics, Chemistry, and Biology}}}}\ (\bibinfo  {publisher}
  {Springer-Verlag},\ \bibinfo {year} {1983})\BibitemShut {NoStop}%
\bibitem [{\citenamefont {Holmes}\ \emph {et~al.}(2012)\citenamefont {Holmes},
  \citenamefont {Lumley}, \citenamefont {Berkooz},\ and\ \citenamefont
  {Rowley}}]{holmes}%
  \BibitemOpen
  \bibfield  {author} {\bibinfo {author} {\bibfnamefont {P.~J.}\ \bibnamefont
  {Holmes}}, \bibinfo {author} {\bibfnamefont {J.~L.}\ \bibnamefont {Lumley}},
  \bibinfo {author} {\bibfnamefont {G.}~\bibnamefont {Berkooz}}, \ and\
  \bibinfo {author} {\bibfnamefont {C.~W.}\ \bibnamefont {Rowley}},\
  }\href@noop {} {\emph {\bibinfo {title} {{Turbulence, Coherent Structures,
  Dynamical Systems and Symmetry. Cambridge Monographs in Mechanics. 2nd
  Ed}}}}\ (\bibinfo  {publisher} {Cambridge University Press},\ \bibinfo {year}
  {2012})\BibitemShut {NoStop}%
\bibitem [{\citenamefont {Mancho}(1996)}]{manchotesis}%
  \BibitemOpen
  \bibfield  {author} {\bibinfo {author} {\bibfnamefont {A.~M.}\ \bibnamefont
  {Mancho}},\ }\href@noop {} {\emph {\bibinfo {title} {{Formaci\'on y
  Din\'amica de Estructuras en Sistemas F\'isicos con Inestabilidades
  Hidrotermales}}}}\ (\bibinfo  {publisher} {Phd Thesis. Universidad de
  Navarra},\ \bibinfo {year} {1996})\BibitemShut {NoStop}%
\bibitem [{\citenamefont {Dawson}\ and\ \citenamefont
  {Mancho}(1997)}]{silvina}%
  \BibitemOpen
  \bibfield  {author} {\bibinfo {author} {\bibfnamefont {S.~P.}\ \bibnamefont
  {Dawson}}\ and\ \bibinfo {author} {\bibfnamefont {A.~M.}\ \bibnamefont
  {Mancho}},\ }\bibfield  {title} {\enquote {\bibinfo {title} {{Collections of
  Heteroclinic Cycles in the Kuramoto-Sivashinsky Equation}},}\ }\href@noop {}
  {\bibfield  {journal} {\bibinfo  {journal} {Physica D}\ }\textbf {\bibinfo
  {volume} {100}},\ \bibinfo {pages} {231--256} (\bibinfo {year}
  {1997})}\BibitemShut {NoStop}%
\bibitem [{\citenamefont {Rowley}\ and\ \citenamefont {Marsden}(2000)}]{row1}%
  \BibitemOpen
  \bibfield  {author} {\bibinfo {author} {\bibfnamefont {C.~W.}\ \bibnamefont
  {Rowley}}\ and\ \bibinfo {author} {\bibfnamefont {J.~E.}\ \bibnamefont
  {Marsden}},\ }\bibfield  {title} {\enquote {\bibinfo {title} {{
  Reconstruction equations and the Karhunen–Loe\'eve expansion for systems
  with symmetry}},}\ }\href@noop {} {\bibfield  {journal} {\bibinfo  {journal}
  {Physica D}\ }\textbf {\bibinfo {volume} {142}},\ \bibinfo {pages} {1--19}
  (\bibinfo {year} {2000})}\BibitemShut {NoStop}%
\bibitem [{\citenamefont {Rowley}(2005)}]{row2}%
  \BibitemOpen
  \bibfield  {author} {\bibinfo {author} {\bibfnamefont {C.~W.}\ \bibnamefont
  {Rowley}},\ }\bibfield  {title} {\enquote {\bibinfo {title} {{ Model
  reduction for fluids using balanced proper orthogonal decomposition}},}\
  }\href@noop {} {\bibfield  {journal} {\bibinfo  {journal} {International
  Journal of Bifurcation and Chaos}\ }\textbf {\bibinfo {volume} {15}},\
  \bibinfo {pages} {997–1013} (\bibinfo {year} {2005})}\BibitemShut {NoStop}%
\bibitem [{\citenamefont {Dovis}\ and\ \citenamefont {Musumeci}(2016)}]{Dovis}%
  \BibitemOpen
  \bibfield  {author} {\bibinfo {author} {\bibfnamefont {F.}~\bibnamefont
  {Dovis}}\ and\ \bibinfo {author} {\bibfnamefont {L.}~\bibnamefont
  {Musumeci}},\ }\bibfield  {title} {\enquote {\bibinfo {title} {Use of the
  karhunen-lo\`{e}ve transform for interference detection and mitigation in
  gnss},}\ }\href {\doibase 10.1016/j.icte.2016.02.008} {\bibfield  {journal}
  {\bibinfo  {journal} {ICT Express}\ }\textbf {\bibinfo {volume} {2}}
  (\bibinfo {year} {2016}),\ 10.1016/j.icte.2016.02.008}\BibitemShut {NoStop}%
\bibitem [{\citenamefont {Musumeci}\ and\ \citenamefont
  {Dovis}(2012)}]{Musumeci}%
  \BibitemOpen
  \bibfield  {author} {\bibinfo {author} {\bibfnamefont {L.}~\bibnamefont
  {Musumeci}}\ and\ \bibinfo {author} {\bibfnamefont {F.}~\bibnamefont
  {Dovis}},\ }\bibfield  {title} {\enquote {\bibinfo {title} {A comparison of
  transformed-domain techniques for pulsed interference removal on gnss
  signals},}\ }\href {\doibase 10.1109/ICL-GNSS.2012.6253131} {\bibfield
  {journal} {\bibinfo  {journal} {International Conference on Localization and
  GNSS}\ } (\bibinfo {year} {2012}),\
  10.1109/ICL-GNSS.2012.6253131}\BibitemShut {NoStop}%
\bibitem [{\citenamefont {Sharifi-Tehrani}, \citenamefont {Sabahi},\ and\
  \citenamefont {Raees~Danaee}(2021)}]{Sharifi}%
  \BibitemOpen
  \bibfield  {author} {\bibinfo {author} {\bibfnamefont {O.}~\bibnamefont
  {Sharifi-Tehrani}}, \bibinfo {author} {\bibfnamefont {M.~F.}\ \bibnamefont
  {Sabahi}}, \ and\ \bibinfo {author} {\bibfnamefont {M.}~\bibnamefont
  {Raees~Danaee}},\ }\bibfield  {title} {\enquote {\bibinfo {title} {Efficient
  gnss jamming mitigation using the marcenkopastur law and karhunen-loeve
  decomposition},}\ }\href {\doibase 10.1109/TAES.2021.3131400} {\bibfield
  {journal} {\bibinfo  {journal} {IEEE Transactions on Aerospace and Electronic
  Systems}\ } (\bibinfo {year} {2021}),\ 10.1109/TAES.2021.3131400}\BibitemShut
  {NoStop}%
\bibitem [{\citenamefont {Liu}(1999)}]{Liu}%
  \BibitemOpen
  \bibfield  {author} {\bibinfo {author} {\bibfnamefont {X.}~\bibnamefont
  {Liu}},\ }\bibfield  {title} {\enquote {\bibinfo {title} {Ground roll
  suppression using the karhunen-loeve transform},}\ }\href {\doibase
  10.1190/1.1444562} {\bibfield  {journal} {\bibinfo  {journal} {Geophysics}\
  }\textbf {\bibinfo {volume} {64}} (\bibinfo {year} {1999}),\
  10.1190/1.1444562}\BibitemShut {NoStop}%
\bibitem [{\citenamefont {Serdyukov}(2022)}]{Serd}%
  \BibitemOpen
  \bibfield  {author} {\bibinfo {author} {\bibfnamefont {A.~S.}\ \bibnamefont
  {Serdyukov}},\ }\bibfield  {title} {\enquote {\bibinfo {title} {Ground-roll
  extraction using the karhunen-loeve transform in the time-frequency
  domain},}\ }\href {\doibase 10.1190/geo2021-0453.1} {\bibfield  {journal}
  {\bibinfo  {journal} {Geophysics}\ }\textbf {\bibinfo {volume} {87}}
  (\bibinfo {year} {2022}),\ 10.1190/geo2021-0453.1}\BibitemShut {NoStop}%
\bibitem [{\citenamefont {Tipireddy}, \citenamefont {Barajas-Solano},\ and\
  \citenamefont {Tartakovsky}(2020)}]{Rama}%
  \BibitemOpen
  \bibfield  {author} {\bibinfo {author} {\bibfnamefont {R.}~\bibnamefont
  {Tipireddy}}, \bibinfo {author} {\bibfnamefont {D.~A.}\ \bibnamefont
  {Barajas-Solano}}, \ and\ \bibinfo {author} {\bibfnamefont {A.~M.}\
  \bibnamefont {Tartakovsky}},\ }\bibfield  {title} {\enquote {\bibinfo {title}
  {Conditional karhunen-loève expansion for uncertainty quantification and
  active learning in partial differential equation models},}\ }\href {\doibase
  10.1016/j.jcp.2020.109604} {\bibfield  {journal} {\bibinfo  {journal}
  {Conditional Karhunen-Loeve expansion for uncertainty quantification and
  active learning in partial differential equation models}\ }\textbf {\bibinfo
  {volume} {418}} (\bibinfo {year} {2020}),\
  10.1016/j.jcp.2020.109604}\BibitemShut {NoStop}%
\bibitem [{\citenamefont {Khonina}, \citenamefont {Volotovsky},\ and\
  \citenamefont {Kirilenko}(2020)}]{Khonina}%
  \BibitemOpen
  \bibfield  {author} {\bibinfo {author} {\bibfnamefont {S.~N.}\ \bibnamefont
  {Khonina}}, \bibinfo {author} {\bibfnamefont {S.~G.}\ \bibnamefont
  {Volotovsky}}, \ and\ \bibinfo {author} {\bibfnamefont {M.~S.}\ \bibnamefont
  {Kirilenko}},\ }\bibfield  {title} {\enquote {\bibinfo {title} {A method of
  generating a random optical field using the karhunen-loeve expansion to
  simulate atmospheric turbulence},}\ }\href {\doibase
  10.18287/2412-6179-CO-680} {\bibfield  {journal} {\bibinfo  {journal}
  {Computer Optics}\ }\textbf {\bibinfo {volume} {44}},\ \bibinfo {pages}
  {53–59} (\bibinfo {year} {2020})}\BibitemShut {NoStop}%
\bibitem [{\citenamefont {Eady}(1949)}]{Eady49}%
  \BibitemOpen
  \bibfield  {author} {\bibinfo {author} {\bibfnamefont {E.~T.}\ \bibnamefont
  {Eady}},\ }\bibfield  {title} {\enquote {\bibinfo {title} {Long waves and
  cyclone waves},}\ }\href {\doibase 10.1111/j.2153-3490.1949.tb01265.x}
  {\bibfield  {journal} {\bibinfo  {journal} {Tellus}\ }\textbf {\bibinfo
  {volume} {1}},\ \bibinfo {pages} {33--52} (\bibinfo {year}
  {1949})}\BibitemShut {NoStop}%
\bibitem [{\citenamefont {Harlander}\ \emph {et~al.}(2012)\citenamefont
  {Harlander}, \citenamefont {Wenzel}, \citenamefont {Alexandrov},
  \citenamefont {Wang},\ and\ \citenamefont {Egbers}}]{Harlander12}%
  \BibitemOpen
  \bibfield  {author} {\bibinfo {author} {\bibfnamefont {U.}~\bibnamefont
  {Harlander}}, \bibinfo {author} {\bibfnamefont {J.}~\bibnamefont {Wenzel}},
  \bibinfo {author} {\bibfnamefont {K.}~\bibnamefont {Alexandrov}}, \bibinfo
  {author} {\bibfnamefont {Y.}~\bibnamefont {Wang}}, \ and\ \bibinfo {author}
  {\bibfnamefont {C.}~\bibnamefont {Egbers}},\ }\bibfield  {title} {\enquote
  {\bibinfo {title} {Simultaneous piv and thermography measurements of
  partially blocked flow in a differentially heated rotating annulus},}\ }\href
  {\doibase 10.1007/s00348-011-1195-y} {\bibfield  {journal} {\bibinfo
  {journal} {Experiments in Fluids}\ }\textbf {\bibinfo {volume} {52}},\
  \bibinfo {pages} {1077--1087} (\bibinfo {year} {2012})}\BibitemShut {NoStop}%
\bibitem [{\citenamefont {Rodda}, \citenamefont {Harlander},\ and\
  \citenamefont {Vincze}(2022)}]{Rodda_etal22}%
  \BibitemOpen
  \bibfield  {author} {\bibinfo {author} {\bibfnamefont {C.}~\bibnamefont
  {Rodda}}, \bibinfo {author} {\bibfnamefont {U.}~\bibnamefont {Harlander}}, \
  and\ \bibinfo {author} {\bibfnamefont {M.}~\bibnamefont {Vincze}},\
  }\bibfield  {title} {\enquote {\bibinfo {title} {{Jet stream variability in a
  polar warming scenario--a laboratory perspective}},}\ }\href {\doibase
  https://wcd.copernicus.org/articles/3/937/2022/} {\bibfield  {journal}
  {\bibinfo  {journal} {Weather Clim. Dynam.}\ }\textbf {\bibinfo {volume}
  {3}},\ \bibinfo {pages} {937--950} (\bibinfo {year} {2022})}\BibitemShut
  {NoStop}%
\bibitem [{\citenamefont {Mancho}, \citenamefont {Small},\ and\ \citenamefont
  {Wiggins}(2006)}]{interp}%
  \BibitemOpen
  \bibfield  {author} {\bibinfo {author} {\bibfnamefont {A.~M.}\ \bibnamefont
  {Mancho}}, \bibinfo {author} {\bibfnamefont {D.}~\bibnamefont {Small}}, \
  and\ \bibinfo {author} {\bibfnamefont {S.}~\bibnamefont {Wiggins}},\
  }\bibfield  {title} {\enquote {\bibinfo {title} {{A comparison of methods for
  interpolating chaotic flows from discrete velocity data}},}\ }\href@noop {}
  {\bibfield  {journal} {\bibinfo  {journal} {Computers \& Fluids}\ }\textbf
  {\bibinfo {volume} {35}},\ \bibinfo {pages} {416--428} (\bibinfo {year}
  {2006})}\BibitemShut {NoStop}%
\bibitem [{\citenamefont {Arnold}(1963)}]{kam1}%
  \BibitemOpen
  \bibfield  {author} {\bibinfo {author} {\bibfnamefont {V.~I.}\ \bibnamefont
  {Arnold}},\ }\bibfield  {title} {\enquote {\bibinfo {title} {{Proof of A. N.
  Kolmogorov’s theorem on the preservation of quasiperodic motions under
  small perturbations of the Hamiltonian }},}\ }\href@noop {} {\bibfield
  {journal} {\bibinfo  {journal} {Russ. Math. Surveys}\ }\textbf {\bibinfo
  {volume} {18}},\ \bibinfo {pages} {9--36} (\bibinfo {year}
  {1963})}\BibitemShut {NoStop}%
\bibitem [{\citenamefont {Kolmogorov}(1954)}]{kam2}%
  \BibitemOpen
  \bibfield  {author} {\bibinfo {author} {\bibfnamefont {A.~N.}\ \bibnamefont
  {Kolmogorov}},\ }\bibfield  {title} {\enquote {\bibinfo {title} {{On
  conservation of conditionally periodic motions under small perturbations of
  the Hamiltonian }},}\ }\href@noop {} {\bibfield  {journal} {\bibinfo
  {journal} {Dokl. Akad. Nauk. USSR}\ }\textbf {\bibinfo {volume} {98}},\
  \bibinfo {pages} {527--530} (\bibinfo {year} {1954})}\BibitemShut {NoStop}%
\bibitem [{\citenamefont {Moser}(1955)}]{kam3}%
  \BibitemOpen
  \bibfield  {author} {\bibinfo {author} {\bibfnamefont {J.}~\bibnamefont
  {Moser}},\ }\bibfield  {title} {\enquote {\bibinfo {title} {{
  Stabilitätsverhalten kanonischer Differentialgleichungssysteme, Nachrichten
  der Akademie der Wissenschaften }},}\ }\href@noop {} {\bibfield  {journal}
  {\bibinfo  {journal} {in Göttingen. II. Mathematisch-Physikalische Klasse}\
  }\textbf {\bibinfo {volume} {98}},\ \bibinfo {pages} {87--120} (\bibinfo
  {year} {1955})}\BibitemShut {NoStop}%
\bibitem [{\citenamefont {Nekhoroshev}(1977)}]{nek}%
  \BibitemOpen
  \bibfield  {author} {\bibinfo {author} {\bibfnamefont {N.~N.}\ \bibnamefont
  {Nekhoroshev}},\ }\bibfield  {title} {\enquote {\bibinfo {title} {{An
  exponential estimate on the time of stability of nearly-integrable
  Hamiltonian systems }},}\ }\href@noop {} {\bibfield  {journal} {\bibinfo
  {journal} {Russ. Math. Surveys}\ }\textbf {\bibinfo {volume} {32}},\ \bibinfo
  {pages} {1--65} (\bibinfo {year} {1977})}\BibitemShut {NoStop}%
\bibitem [{\citenamefont {Velasco~Fuentes}(1994)}]{velasco}%
  \BibitemOpen
  \bibfield  {author} {\bibinfo {author} {\bibfnamefont {O.~U.}\ \bibnamefont
  {Velasco~Fuentes}},\ }\emph {\bibinfo {title} {{Two-Dimensional Vortices with
  Background Vorticity}}},\ \href@noop {} {Ph.D. thesis},\ \bibinfo  {school}
  {Eindhoven University of Technology} (\bibinfo {year} {1994})\BibitemShut
  {NoStop}%
\bibitem [{\citenamefont {Mendoza}\ and\ \citenamefont
  {Mancho}(2010)}]{Mendoza}%
  \BibitemOpen
  \bibfield  {author} {\bibinfo {author} {\bibfnamefont {C.}~\bibnamefont
  {Mendoza}}\ and\ \bibinfo {author} {\bibfnamefont {A.~M.}\ \bibnamefont
  {Mancho}},\ }\bibfield  {title} {\enquote {\bibinfo {title} {The hidden
  geometry of ocean flows},}\ }\href {\doibase 10.1103/PhysRevLett.105.038501}
  {\bibfield  {journal} {\bibinfo  {journal} {Physical Review Letters}\
  }\textbf {\bibinfo {volume} {105}} (\bibinfo {year} {2010}),\
  10.1103/PhysRevLett.105.038501}\BibitemShut {NoStop}%
\bibitem [{\citenamefont {Mancho}\ \emph {et~al.}(2013)\citenamefont {Mancho},
  \citenamefont {Wiggins}, \citenamefont {Curbelo},\ and\ \citenamefont
  {Mendoza}}]{cnsns}%
  \BibitemOpen
  \bibfield  {author} {\bibinfo {author} {\bibfnamefont {A.~M.}\ \bibnamefont
  {Mancho}}, \bibinfo {author} {\bibfnamefont {S.}~\bibnamefont {Wiggins}},
  \bibinfo {author} {\bibfnamefont {J.}~\bibnamefont {Curbelo}}, \ and\
  \bibinfo {author} {\bibfnamefont {C.}~\bibnamefont {Mendoza}},\ }\bibfield
  {title} {\enquote {\bibinfo {title} {{Lagrangian Descriptors: A Method for
  Revealing Phase Space Structures of General Time Dependent Dynamical
  Systems}},}\ }\href {\doibase 10.1016/j.cnsns.2013.05.002} {\bibfield
  {journal} {\bibinfo  {journal} {Communications in Nonlinear Science and
  Numerical Simulation}\ }\textbf {\bibinfo {volume} {18}},\ \bibinfo {pages}
  {3530--3557} (\bibinfo {year} {2013})}\BibitemShut {NoStop}%
\bibitem [{\citenamefont {Lopesino}\ \emph {et~al.}(2015)\citenamefont
  {Lopesino}, \citenamefont {Balibrea}, \citenamefont {Wiggins},\ and\
  \citenamefont {Mancho}}]{carlos}%
  \BibitemOpen
  \bibfield  {author} {\bibinfo {author} {\bibfnamefont {C.}~\bibnamefont
  {Lopesino}}, \bibinfo {author} {\bibfnamefont {F.}~\bibnamefont {Balibrea}},
  \bibinfo {author} {\bibfnamefont {S.}~\bibnamefont {Wiggins}}, \ and\
  \bibinfo {author} {\bibfnamefont {A.~M.}\ \bibnamefont {Mancho}},\ }\bibfield
   {title} {\enquote {\bibinfo {title} {{Lagrangian Descriptors for Two
  Dimensional, Area Preserving Autonomous and Nonautonomous Maps}},}\ }\href
  {\doibase 10.1016/j.cnsns.2015.02.022} {\bibfield  {journal} {\bibinfo
  {journal} {Communications in Nonlinear Science and Numerical Simulation}\
  }\textbf {\bibinfo {volume} {27}},\ \bibinfo {pages} {40--51} (\bibinfo
  {year} {2015})}\BibitemShut {NoStop}%
\bibitem [{\citenamefont {Lopesino}\ \emph {et~al.}(2017)\citenamefont
  {Lopesino}, \citenamefont {Balibrea-Iniesta}, \citenamefont
  {Garc\'ia-Garrido}, \citenamefont {Wiggins},\ and\ \citenamefont
  {Mancho}}]{lopesino2017}%
  \BibitemOpen
  \bibfield  {author} {\bibinfo {author} {\bibfnamefont {C.}~\bibnamefont
  {Lopesino}}, \bibinfo {author} {\bibfnamefont {F.}~\bibnamefont
  {Balibrea-Iniesta}}, \bibinfo {author} {\bibfnamefont {V.~J.}\ \bibnamefont
  {Garc\'ia-Garrido}}, \bibinfo {author} {\bibfnamefont {S.}~\bibnamefont
  {Wiggins}}, \ and\ \bibinfo {author} {\bibfnamefont {A.~M.}\ \bibnamefont
  {Mancho}},\ }\bibfield  {title} {\enquote {\bibinfo {title} {A theoretical
  framework for lagrangian descriptors},}\ }\href {\doibase
  10.1142/S0218127417300014} {\bibfield  {journal} {\bibinfo  {journal}
  {International Journal of Bifurcation and Chaos}\ }\textbf {\bibinfo {volume}
  {27}},\ \bibinfo {pages} {1730001} (\bibinfo {year} {2017})}\BibitemShut
  {NoStop}%
\bibitem [{\citenamefont {Garc\'ia-Garrido}\ \emph {et~al.}(2018)\citenamefont
  {Garc\'ia-Garrido}, \citenamefont {Balibrea-Iniesta}, \citenamefont
  {Wiggins}, \citenamefont {Mancho},\ and\ \citenamefont {Lopesino}}]{gg2018b}%
  \BibitemOpen
  \bibfield  {author} {\bibinfo {author} {\bibfnamefont {V.~J.}\ \bibnamefont
  {Garc\'ia-Garrido}}, \bibinfo {author} {\bibfnamefont {F.}~\bibnamefont
  {Balibrea-Iniesta}}, \bibinfo {author} {\bibfnamefont {S.}~\bibnamefont
  {Wiggins}}, \bibinfo {author} {\bibfnamefont {A.~M.}\ \bibnamefont {Mancho}},
  \ and\ \bibinfo {author} {\bibfnamefont {C.}~\bibnamefont {Lopesino}},\
  }\bibfield  {title} {\enquote {\bibinfo {title} {Detection of phase space
  structures of the cat map with lagrangian descriptors},}\ }\href@noop {}
  {\bibfield  {journal} {\bibinfo  {journal} {Regular and Chaotic Dynamics}\
  }\textbf {\bibinfo {volume} {23}},\ \bibinfo {pages} {751--766} (\bibinfo
  {year} {2018})}\BibitemShut {NoStop}%
\bibitem [{\citenamefont {de~la C{\'a}mara}\ \emph
  {et~al.}(2012{\natexlab{b}})\citenamefont {de~la C{\'a}mara}, \citenamefont
  {Mancho}, \citenamefont {Ide}, \citenamefont {Mechoso},\ and\ \citenamefont
  {Serrano}}]{delacamara2012}%
  \BibitemOpen
  \bibfield  {author} {\bibinfo {author} {\bibfnamefont {A.}~\bibnamefont
  {de~la C{\'a}mara}}, \bibinfo {author} {\bibfnamefont {A.~M.}\ \bibnamefont
  {Mancho}}, \bibinfo {author} {\bibfnamefont {K.}~\bibnamefont {Ide}},
  \bibinfo {author} {\bibfnamefont {C.~R.}\ \bibnamefont {Mechoso}}, \ and\
  \bibinfo {author} {\bibfnamefont {E.}~\bibnamefont {Serrano}},\ }\bibfield
  {title} {\enquote {\bibinfo {title} {{Routes of transport across the
  Antarctic polar vortex in the southern spring}},}\ }\href {\doibase
  10.1175/JAS-D-11-0142.1} {\bibfield  {journal} {\bibinfo  {journal} {J.
  Atmos. Sci.}\ }\textbf {\bibinfo {volume} {69}},\ \bibinfo {pages} {753--767}
  (\bibinfo {year} {2012}{\natexlab{b}})}\BibitemShut {NoStop}%
\bibitem [{\citenamefont {Garc\'{i}a-Garrido}\ \emph
  {et~al.}(2016)\citenamefont {Garc\'{i}a-Garrido}, \citenamefont {Ramos},
  \citenamefont {Mancho}, \citenamefont {Coca},\ and\ \citenamefont
  {Wiggins}}]{ggrmcw16}%
  \BibitemOpen
  \bibfield  {author} {\bibinfo {author} {\bibfnamefont {V.~J.}\ \bibnamefont
  {Garc\'{i}a-Garrido}}, \bibinfo {author} {\bibfnamefont {A.}~\bibnamefont
  {Ramos}}, \bibinfo {author} {\bibfnamefont {A.~M.}\ \bibnamefont {Mancho}},
  \bibinfo {author} {\bibfnamefont {J.}~\bibnamefont {Coca}}, \ and\ \bibinfo
  {author} {\bibfnamefont {S.}~\bibnamefont {Wiggins}},\ }\bibfield  {title}
  {\enquote {\bibinfo {title} {A dynamical systems perspective for a real-time
  response to a marine oil spill},}\ }\href {\doibase
  https://doi.org/10.1016/j.marpolbul.2016.08.018} {\bibfield  {journal}
  {\bibinfo  {journal} {Marine Pollution Bulletin}\ }\textbf {\bibinfo {volume}
  {112}},\ \bibinfo {pages} {201--210} (\bibinfo {year} {2016})}\BibitemShut
  {NoStop}%
\bibitem [{\citenamefont {Ramos}\ \emph {et~al.}(2018)\citenamefont {Ramos},
  \citenamefont {Garc{\'i}a-Garrido}, \citenamefont {Mancho}, \citenamefont
  {Wiggins}, \citenamefont {Coca}, \citenamefont {Glenn}, \citenamefont
  {Schofield}, \citenamefont {Kohut}, \citenamefont {Aragon}, \citenamefont
  {Kerfoot}, \citenamefont {Haskins}, \citenamefont {Miles}, \citenamefont
  {Haldeman}, \citenamefont {Strandskov}, \citenamefont {Allsup}, \citenamefont
  {Jones},\ and\ \citenamefont {Shapiro}}]{ramos2018}%
  \BibitemOpen
  \bibfield  {author} {\bibinfo {author} {\bibfnamefont {A.~G.}\ \bibnamefont
  {Ramos}}, \bibinfo {author} {\bibfnamefont {V.~J.}\ \bibnamefont
  {Garc{\'i}a-Garrido}}, \bibinfo {author} {\bibfnamefont {A.~M.}\ \bibnamefont
  {Mancho}}, \bibinfo {author} {\bibfnamefont {S.}~\bibnamefont {Wiggins}},
  \bibinfo {author} {\bibfnamefont {J.}~\bibnamefont {Coca}}, \bibinfo {author}
  {\bibfnamefont {S.}~\bibnamefont {Glenn}}, \bibinfo {author} {\bibfnamefont
  {O.}~\bibnamefont {Schofield}}, \bibinfo {author} {\bibfnamefont
  {J.}~\bibnamefont {Kohut}}, \bibinfo {author} {\bibfnamefont
  {D.}~\bibnamefont {Aragon}}, \bibinfo {author} {\bibfnamefont
  {J.}~\bibnamefont {Kerfoot}}, \bibinfo {author} {\bibfnamefont
  {T.}~\bibnamefont {Haskins}}, \bibinfo {author} {\bibfnamefont
  {T.}~\bibnamefont {Miles}}, \bibinfo {author} {\bibfnamefont
  {C.}~\bibnamefont {Haldeman}}, \bibinfo {author} {\bibfnamefont
  {N.}~\bibnamefont {Strandskov}}, \bibinfo {author} {\bibfnamefont
  {B.}~\bibnamefont {Allsup}}, \bibinfo {author} {\bibfnamefont
  {C.}~\bibnamefont {Jones}}, \ and\ \bibinfo {author} {\bibfnamefont
  {J.}~\bibnamefont {Shapiro}},\ }\bibfield  {title} {\enquote {\bibinfo
  {title} {Lagrangian coherent structure assisted path planning for
  transoceanic autonomous underwater vehicle missions.}}\ }\href {\doibase
  10.1038/s41598-018-23028-8} {\bibfield  {journal} {\bibinfo  {journal}
  {Scientific Reports}\ }\textbf {\bibinfo {volume} {8}},\ \bibinfo {pages}
  {4575} (\bibinfo {year} {2018})}\BibitemShut {NoStop}%
\bibitem [{\citenamefont {Garc{\'\i}a-S{\'a}nchez}\ \emph
  {et~al.}(2021)\citenamefont {Garc{\'\i}a-S{\'a}nchez}, \citenamefont
  {Mancho}, \citenamefont {Ramos}, \citenamefont {Coca}, \citenamefont
  {P{\'e}rez-G{\'o}mez}, \citenamefont {{\'A}lvarez-Fanjul}, \citenamefont
  {Sotillo}, \citenamefont {Garc{\'\i}a-Le{\'o}n}, \citenamefont
  {Garc\'{i}a-Garrido},\ and\ \citenamefont {Wiggins}}]{garciasanchez2020}%
  \BibitemOpen
  \bibfield  {author} {\bibinfo {author} {\bibfnamefont {G.}~\bibnamefont
  {Garc{\'\i}a-S{\'a}nchez}}, \bibinfo {author} {\bibfnamefont {A.~M.}\
  \bibnamefont {Mancho}}, \bibinfo {author} {\bibfnamefont {A.~G.}\
  \bibnamefont {Ramos}}, \bibinfo {author} {\bibfnamefont {J.}~\bibnamefont
  {Coca}}, \bibinfo {author} {\bibfnamefont {B.}~\bibnamefont
  {P{\'e}rez-G{\'o}mez}}, \bibinfo {author} {\bibfnamefont {E.}~\bibnamefont
  {{\'A}lvarez-Fanjul}}, \bibinfo {author} {\bibfnamefont {M.~G.}\ \bibnamefont
  {Sotillo}}, \bibinfo {author} {\bibfnamefont {M.}~\bibnamefont
  {Garc{\'\i}a-Le{\'o}n}}, \bibinfo {author} {\bibfnamefont {V.~J.}\
  \bibnamefont {Garc\'{i}a-Garrido}}, \ and\ \bibinfo {author} {\bibfnamefont
  {S.}~\bibnamefont {Wiggins}},\ }\bibfield  {title} {\enquote {\bibinfo
  {title} {Very high resolution tools for the monitoring and assessment of
  environmental hazards in coastal areas},}\ }\href@noop {} {\bibfield
  {journal} {\bibinfo  {journal} {Frontiers in Marine Science}\ }\textbf
  {\bibinfo {volume} {7}} (\bibinfo {year} {2021})}\BibitemShut {NoStop}%
\bibitem [{\citenamefont {Craven}\ and\ \citenamefont
  {Hernandez}(0016)}]{Craven}%
  \BibitemOpen
  \bibfield  {author} {\bibinfo {author} {\bibfnamefont {G.~T.}\ \bibnamefont
  {Craven}}\ and\ \bibinfo {author} {\bibfnamefont {R.}~\bibnamefont
  {Hernandez}},\ }\bibfield  {title} {\enquote {\bibinfo {title} {{
  Deconstructing field-induced ketene isomerization through Lagrangian
  descriptors}},}\ }\href@noop {} {\bibfield  {journal} {\bibinfo  {journal}
  {Phys. Chem. Chem. Phys.}\ }\textbf {\bibinfo {volume} {18}},\ \bibinfo
  {pages} {4008} (\bibinfo {year} {20016})}\BibitemShut {NoStop}%
\bibitem [{\citenamefont {Agaoglou}\ \emph {et~al.}(2021)\citenamefont
  {Agaoglou}, \citenamefont {Garc\'{i}a-Garrido}, \citenamefont {Katsanikas},\
  and\ \citenamefont {Wiggins}}]{Agaoglou1}%
  \BibitemOpen
  \bibfield  {author} {\bibinfo {author} {\bibfnamefont {M.}~\bibnamefont
  {Agaoglou}}, \bibinfo {author} {\bibfnamefont {V.~J.}\ \bibnamefont
  {Garc\'{i}a-Garrido}}, \bibinfo {author} {\bibfnamefont {M.}~\bibnamefont
  {Katsanikas}}, \ and\ \bibinfo {author} {\bibfnamefont {S.}~\bibnamefont
  {Wiggins}},\ }\bibfield  {title} {\enquote {\bibinfo {title} {Visualizing the
  phase space of the hei$_2$ van der waals complex using lagrangian
  descriptors},}\ }\href {\doibase 10.1016/j.cnsns.2021.105993} {\bibfield
  {journal} {\bibinfo  {journal} {Communications in Nonlinear Science and
  Numerical Simulation}\ }\textbf {\bibinfo {volume} {103}} (\bibinfo {year}
  {2021}),\ 10.1016/j.cnsns.2021.105993}\BibitemShut {NoStop}%
\bibitem [{\citenamefont {Garc\'{i}a-Garrido}, \citenamefont {Agaoglou},\ and\
  \citenamefont {Wiggins}(2020)}]{Garcia2}%
  \BibitemOpen
  \bibfield  {author} {\bibinfo {author} {\bibfnamefont {V.~J.}\ \bibnamefont
  {Garc\'{i}a-Garrido}}, \bibinfo {author} {\bibfnamefont {M.}~\bibnamefont
  {Agaoglou}}, \ and\ \bibinfo {author} {\bibfnamefont {S.}~\bibnamefont
  {Wiggins}},\ }\bibfield  {title} {\enquote {\bibinfo {title} {Exploring
  isomerization dynamics on a potential energy surface with an index-2 saddle
  using lagrangian descriptors},}\ }\href {\doibase
  10.1016/j.cnsns.2020.105331} {\bibfield  {journal} {\bibinfo  {journal}
  {Communications in Nonlinear Science and Numerical Simulation}\ }\textbf
  {\bibinfo {volume} {89}} (\bibinfo {year} {2020}),\
  10.1016/j.cnsns.2020.105331}\BibitemShut {NoStop}%
\bibitem [{\citenamefont {Agaoglou}\ \emph {et~al.}(2019)\citenamefont
  {Agaoglou}, \citenamefont {Aguilar-Sanjuan}, \citenamefont
  {Garc{\'\i}a-Garrido}, \citenamefont {Garc{\'\i}a-Meseguer}, \citenamefont
  {Gonz{\'a}lez-Montoya}, \citenamefont {Katsanikas}, \citenamefont
  {Kraj{\v{n}}{\'a}k}, \citenamefont {Naik},\ and\ \citenamefont
  {Wiggins}}]{Agaoglou2}%
  \BibitemOpen
  \bibfield  {author} {\bibinfo {author} {\bibfnamefont {M.}~\bibnamefont
  {Agaoglou}}, \bibinfo {author} {\bibfnamefont {B.}~\bibnamefont
  {Aguilar-Sanjuan}}, \bibinfo {author} {\bibfnamefont {V.}~\bibnamefont
  {Garc{\'\i}a-Garrido}}, \bibinfo {author} {\bibfnamefont {R.}~\bibnamefont
  {Garc{\'\i}a-Meseguer}}, \bibinfo {author} {\bibfnamefont {F.}~\bibnamefont
  {Gonz{\'a}lez-Montoya}}, \bibinfo {author} {\bibfnamefont {M.}~\bibnamefont
  {Katsanikas}}, \bibinfo {author} {\bibfnamefont {V.}~\bibnamefont
  {Kraj{\v{n}}{\'a}k}}, \bibinfo {author} {\bibfnamefont {S.}~\bibnamefont
  {Naik}}, \ and\ \bibinfo {author} {\bibfnamefont {S.}~\bibnamefont
  {Wiggins}},\ }\bibfield  {title} {\enquote {\bibinfo {title} {Chemical
  reactions: {A} journey into phase space},}\ }\href {\doibase
  10.5281/zenodo.3568210} {\bibfield  {journal} {\bibinfo  {journal} {Zenodo}\
  } (\bibinfo {year} {2019}),\ 10.5281/zenodo.3568210}\BibitemShut {NoStop}%
\bibitem [{\citenamefont {Agaoglou}\ \emph {et~al.}(2020)\citenamefont
  {Agaoglou}, \citenamefont {Aguilar-Sanjuan}, \citenamefont
  {Garc{\'\i}a-Garrido}, \citenamefont {Gonz{\'a}lez-Montoya}, \citenamefont
  {Katsanikas}, \citenamefont {Kraj{\v{n}}{\'a}k}, \citenamefont {Naik},\ and\
  \citenamefont {Wiggins}}]{Agaoglou3}%
  \BibitemOpen
  \bibfield  {author} {\bibinfo {author} {\bibfnamefont {M.}~\bibnamefont
  {Agaoglou}}, \bibinfo {author} {\bibfnamefont {B.}~\bibnamefont
  {Aguilar-Sanjuan}}, \bibinfo {author} {\bibfnamefont {V.}~\bibnamefont
  {Garc{\'\i}a-Garrido}}, \bibinfo {author} {\bibfnamefont {F.}~\bibnamefont
  {Gonz{\'a}lez-Montoya}}, \bibinfo {author} {\bibfnamefont {M.}~\bibnamefont
  {Katsanikas}}, \bibinfo {author} {\bibfnamefont {V.}~\bibnamefont
  {Kraj{\v{n}}{\'a}k}}, \bibinfo {author} {\bibfnamefont {S.}~\bibnamefont
  {Naik}}, \ and\ \bibinfo {author} {\bibfnamefont {S.}~\bibnamefont
  {Wiggins}},\ }\bibfield  {title} {\enquote {\bibinfo {title} {{Lagrangian
  {D}escriptors: {D}iscovery and {Q}uantification of {P}hase {S}pace
  {S}tructure and {T}ransport}},}\ }\href {\doibase 10.5281/zenodo.3958985}
  {\bibfield  {journal} {\bibinfo  {journal} {Zenodo}\ } (\bibinfo {year}
  {2020}),\ 10.5281/zenodo.3958985}\BibitemShut {NoStop}%
\bibitem [{\citenamefont {Madrid}\ and\ \citenamefont {Mancho}(2009)}]{Madrid}%
  \BibitemOpen
  \bibfield  {author} {\bibinfo {author} {\bibfnamefont {J.~A.~J.}\
  \bibnamefont {Madrid}}\ and\ \bibinfo {author} {\bibfnamefont {A.~M.}\
  \bibnamefont {Mancho}},\ }\bibfield  {title} {\enquote {\bibinfo {title}
  {Distinguished trajectories in time dependent vector fields},}\ }\href
  {\doibase 10.1063/1.3056050} {\bibfield  {journal} {\bibinfo  {journal}
  {Chaos}\ }\textbf {\bibinfo {volume} {19}},\ \bibinfo {pages} {013111}
  (\bibinfo {year} {2009})}\BibitemShut {NoStop}%
\bibitem [{\citenamefont {Provenzale}(1999)}]{Provenzale:1999}%
  \BibitemOpen
  \bibfield  {author} {\bibinfo {author} {\bibfnamefont {A.}~\bibnamefont
  {Provenzale}},\ }\bibfield  {title} {\enquote {\bibinfo {title} {{Transport
  by coherent barotropic vortices}},}\ }\href@noop {} {\bibfield  {journal}
  {\bibinfo  {journal} {Annu. Rev. Fluid Mech.}\ }\textbf {\bibinfo {volume}
  {31}},\ \bibinfo {pages} {55--93} (\bibinfo {year} {1999})}\BibitemShut
  {NoStop}%
\bibitem [{\citenamefont {McIntyre}(1989)}]{McIntyre:1989}%
  \BibitemOpen
  \bibfield  {author} {\bibinfo {author} {\bibfnamefont {M.~E.}\ \bibnamefont
  {McIntyre}},\ }\bibfield  {title} {\enquote {\bibinfo {title} {{On the
  Antarctic ozone hole}},}\ }\href@noop {} {\bibfield  {journal} {\bibinfo
  {journal} {J. Atmos. Terr. Phys.}\ }\textbf {\bibinfo {volume} {51}},\
  \bibinfo {pages} {29--43} (\bibinfo {year} {1989})}\BibitemShut {NoStop}%
\bibitem [{\citenamefont {LaCasce}\ and\ \citenamefont
  {Mahadevan}(2006)}]{LaCasce:2006}%
  \BibitemOpen
  \bibfield  {author} {\bibinfo {author} {\bibfnamefont {J.~H.}\ \bibnamefont
  {LaCasce}}\ and\ \bibinfo {author} {\bibfnamefont {A.}~\bibnamefont
  {Mahadevan}},\ }\bibfield  {title} {\enquote {\bibinfo {title} {Estimating
  subsurface horizontal and vertical velocities from sea-surface
  temperature},}\ }\href {\doibase doi:10.1357/002224006779367267} {\bibfield
  {journal} {\bibinfo  {journal} {Journal of Marine Research}\ }\textbf
  {\bibinfo {volume} {64}},\ \bibinfo {pages} {695--721} (\bibinfo {year}
  {2006})}\BibitemShut {NoStop}%
\end{thebibliography}%

\end{document}